\documentclass[journal]{IEEEtran}

\usepackage{bm}
\usepackage[cmex10]{amsmath}
\usepackage{amssymb}
\usepackage{amsfonts}
\usepackage{amsthm}
\usepackage{array}
\usepackage{dsfont}
\usepackage{graphicx}
\usepackage{epstopdf}
\usepackage{cite}
\usepackage{bbm}
\usepackage{eurosym}
\usepackage{multirow}
\usepackage{mathtools}
\usepackage{color}
\usepackage{caption}
\usepackage{subcaption}
\usepackage{dblfloatfix}    
\usepackage[hyphens]{url}
\usepackage{hyperref}
\usepackage[hyphenbreaks]{breakurl}

\newtheorem{remark}{Remark}

\begin{document}

\title{Privacy-Cost Trade-offs\\ in Smart Electricity Metering Systems}

\author{
    \IEEEauthorblockN{Giulio Giaconi\IEEEauthorrefmark{1}, Deniz G\"{u}nd\"{u}z\IEEEauthorrefmark{1}, H. Vincent Poor\IEEEauthorrefmark{2}}\\
    \IEEEauthorblockA{\IEEEauthorrefmark{1}Department of Electrical and Electronic Engineering, Imperial College London, London,  SW7 2AZ, UK}\\
    \IEEEauthorblockA{\IEEEauthorrefmark{2}Department of Electrical Engineering, Princeton University, Princeton, NJ 08544, USA}
}

\maketitle

\begin{abstract}
Trade-offs between privacy and cost are studied for a smart grid consumer, whose electricity consumption is monitored in almost real time by the utility provider (UP) through smart meter (SM) readings. It is assumed that an electrical battery is available to the consumer, which can be utilized both to achieve privacy and to reduce the energy cost by demand shaping. Privacy is measured via the mean squared distance between the SM readings and a target load profile, while time-of-use (ToU) pricing is considered to compute the cost incurred. The consumer can also sell electricity back to the UP to further improve the privacy-cost trade-off. Two privacy-preserving energy management policies (EMPs) are proposed, which differ in the way the target load profile is characterized. A more practical EMP, which optimizes the energy management less frequently, is also considered. Numerical results are presented to compare the privacy-cost trade-off of these EMPs, considering various privacy indicators.
\end{abstract}

\IEEEpeerreviewmaketitle

\section{Introduction}

Smart meters (SMs) are pivotal components of the smart grid, because they enable two-way communication between each household and the utility provider (UP), the entity that sells energy to consumers. Benefits include having more accurate electricity bills, detecting energy theft and outages faster, introducing time-of-use (ToU) tariffs to match demand with available resources, integrating microgeneration systems, e.g., photovoltaic panels and wind farms, and residential energy storage solutions, and the possibility for the consumers to sell energy to the grid. For these reasons, the SM roll-out is proceeding rapidly and is attracting considerable investments. However, an SM's ability to monitor a user's electricity consumption in almost real-time entails serious implications about consumer privacy. In fact, non-intrusive appliance load monitoring techniques are able to distinguish the power signatures of specific appliances from aggregated household SM measurements, revealing sensitive information about a consumer's life, such as her presence at home, religious beliefs and disabilities \cite{Quinn:2009,Rouf:2012}. SM privacy is also critical for businesses, e.g., factories and data centers, as power consumption data may reveal information about the state of their businesses.

\subsection{Privacy-Aware SM Techniques}

Privacy-preserving methods for SMs can be classified into two families. Firstly, the \textit{smart meter data manipulation} (SMDM) family \cite{Giaconi:2018}, encompasses methods that modify SM measurements before reporting them to the UP, and includes \textit{data obfuscation} \cite{Kim:2011}, \textit{aggregation} \cite{Bohli:2010}, \textit{anonymization} \cite{Petrlic:2010}, \textit{down-sampling} \cite{Cardenas:2012}, and \textit{data-sharing prevention} \cite{Molina:2010} approaches. However, these techniques suffer from several shortcomings \cite{Giaconi:2018}. First, obfuscation approaches add noise to the SM readings, causing a mismatch between the reported values and the real energy consumption, which prevents distribution systems operators (DSOs), i.e., the entities that operate and manage the grid, and UPs from accurately monitoring the grid state. Second, anonymization and aggregation techniques that include the presence of a trusted third party (TTP) only shift the problem of trust from one entity (UP/DSO) to another (TTP). Third, DSOs, UPs, or more generally any eavesdropper can embed additional sensors right outside a household or a business to monitor the energy consumption, without fully relying on SM readings. The second family of privacy-preserving approaches, called the \textit{user demand shaping} (UDS) family \cite{Giaconi:2018}, overcomes these issues by modifying the consumer's actual electricity consumption, called the \textit{user load}, rather than the data sent to the UP. This is achieved by exploiting physical resources, e.g., rechargeable batteries (RBs) or renewable energy sources (RESs), making the user load as different as possible from the SM measurements, called the \textit{grid load} \cite{Kalogridis:2010SGC,Tan:2017TIFS,Giaconi:2017TIFS}. However, UDS techniques may require an initial investment by the user as physical resources need to be installed at the user's premises. Using RBs for privacy preservation may also lead to a quicker physical degradation of RBs \cite{Avula:2018}. Moreover, cost of energy may increase when providing privacy via some UDS approaches. Hence, it is important to consider these aspects when considering UDS techniques for privacy preservation.

In this paper, we adopt UDS techniques because they report the energy taken from the grid accurately, and employ physical resources, e.g., RBs and RESs, which are becoming increasingly available to the consumers. Our aim is to jointly minimize the information leaked about a user and the cost of electricity. While a widely accepted definition of privacy is elusive, privacy is achieved when it is not possible to distinguish a specific appliance load from the aggregated household energy consumption \cite{Kalogridis:2010SGC}. Statistical techniques measure privacy loss by the mutual information between the user and the grid loads \cite{Giaconi:2016,Tan:2017TIFS,Giaconi:2017TIFS,Chin:2017TSG,Yao:2015TSG,Gomez:2015TIFS,Li:2018}, or by computing approximations of it \cite{Chin:2018}; however, this requires the knowledge of the underlying statistics, and the results are typically valid under various simplifications, e.g., assuming independent and identically distributed user load, and over sufficiently long time horizons. An alternative approach is based on the idea that a high degree of privacy can be achieved by flattening the power consumption around a \textit{target load profile}, e.g., minimizing the distance from a completely private profile \cite{Tan:2017TIFS}, \cite{Yang:2014INFOCOM,Yang:2015TSG,Giaconi:2017SGC}. The target load profile can be set to be a constant value over time, typically equal to the average consumption \cite{Tan:2017TIFS, Yang:2014INFOCOM,Yang:2015TSG}. In this model, it is assumed that the energy management unit (EMU), i.e., the system that implements the privacy-preserving energy management policy (EMP) at the user's premises, knows, or, accurately predicts, the load profile for the time period of interest, and obtains the optimal EMP by solving an optimization problem. On the other hand, a completely constant consumption may not be practically viable or desirable, since the energy cost may vary greatly during the system operation due to ToU tariffs. Hence, in \cite{Giaconi:2017SGC} the EMU is allowed to target a different fixed power value for each price period. The flexibility of the latter approach leads to a better overall privacy-cost trade-off; however, such a piecewise constant target profile implies also an inherent information leakage compared to a constant target profile. We follow up on \cite{Tan:2017TIFS} and \cite{Yang:2014INFOCOM,Yang:2015TSG,Giaconi:2017SGC}, and measure the privacy leakage as the squared distance between the grid and target load profiles; however, differently from those works, we consider a more general target load profile, and assume that the consumer has only a partial knowledge of her future energy consumption and energy cost. We note that privacy leakage may be measured as the distance between grid and user loads. However, if the aim is to increase such distance, this might lead to a potentially deterministic strategy for the EMU, e.g., produce a low grid load when the user load is high and vice-versa. Since we also assume that the UP knows the optimal strategy implemented by the EMU, such a strategy would result in a better estimate of the user load by the UP. On the contrary, trying to match the grid load to a specific target profile would make it harder to estimate the user load, e.g., flattening the grid load independent of the user load reveals only the average energy consumption.

The main contributions of this paper are as follows:
\begin{enumerate}
\item While full information about the future electricity consumption is assumed to be available at the EMU in \cite{Tan:2017TIFS} and \cite{Giaconi:2017SGC}, which we call the \textit{long horizon} model (LHM), here we consider a more realistic scenario whereby the consumer's future consumption profile is only partially known to the EMU in a \textit{receding horizon} manner, which we call the \textit{short horizon} model (SHM). The optimal solution at any time is computed only based on the currently available information within the \textit{prediction horizon} by adopting a model predictive controller, recently implemented in an SM setting in \cite{Chin:2017TSG}. We present a detailed comparison of the results for SHM and LHM.
\item We introduce a target load profile computed as a low-pass filtered version of the user load, since higher-frequency components of a user's consumption profile leak more information about her behaviour, compared to lower-frequency components. To the authors' knowledge, this is the first time that such a target profile has been studied in the SM privacy-preservation literature.
\item We propose a more practical EMP that updates the optimal strategy less frequently. The optimal solution is computed in batch, reducing the computational load at the expense of the privacy-cost trade-off. Finally, we compare the privacy-cost trade-offs for all the schemes using real consumption and pricing data.
\end{enumerate}
   
The remainder of this paper is organized as follows. In Section \ref{sec:SystemModel} we present the system model, while in Sections \ref{sec:constantTargetLoad} and \ref{sec:filterTargetLoad} we consider the SHM for a constant and a filtered target load profile, respectively. A more practical EMP with less regular policy updates is analyzed in Section \ref{sec:suboptimal}, while conclusions are drawn in Section \ref{sec:conclusion}.

\subsection{Notation}
For integers $0 < a < b$, $U_{a}^{b}$ denotes the sequence $[U_a,U_{a+1},\ldots,U_b]$, while $U^b \triangleq U_{1}^b$. The positive part $[x]^+$ is equal to $x$ if $x>0$, and $0$ otherwise. When solving optimization problems, we denote the optimal value of a variable with a star, e.g., $G^*$ denotes the optimal value of the parameter $G$.


\section{System Model}\label{sec:SystemModel}

\begin{figure}[!t]
\centering
\includegraphics[width=1\columnwidth]{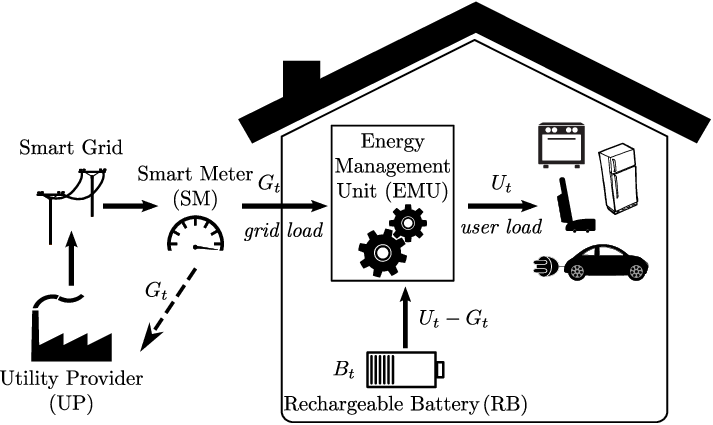}
\caption{The system model. $U_t$, $G_t$ and $U_t-G_t$ are the user load, the grid load, and the energy drawn from the RB at time $t$, respectively. The dashed line represents the meter readings being accurately reported to the UP.}
\label{fig:SystemModel}
\end{figure}

We consider the discrete time system depicted in Fig. \ref{fig:SystemModel}, where $t$ represents one time slot (TS) of duration $D$ seconds, for $1 \leq t \leq N$, where $N$ is the time horizon of interest. For TS $t$, the user load, i.e., the total power requested by all the household appliances within TS $t$, is denoted by $U_t \in \mathcal{U}$, while the grid load is $G_t \in \mathcal{G}$. We remark that the TSs in our model correspond to time instants when the electricity is actually requested by the user and drawn from the grid, rather than the typically longer sampling interval used for sending SM measurements to the UP. We assume that the SM measures and records the output power values at each TS because our aim is to protect consumers' privacy not only from the UP, but also from the DSO or any other attacker that may deploy a sensor on the consumer's power line recording the electricity consumption in almost real-time. An RB of capacity $B_{\max}$ is installed at the user's premises, which is used to both filter the user load to provide privacy, and to shift energy intake from the grid to minimize electricity costs. The EMU computes the amount of energy to draw from the grid, $G_t$, and to exchange with the RB, $U_t - G_t$. Let $B_t\in [0, B_{\max}]$ denote the amount of energy in the RB at the end of TS $t$, and we set $B_0=0$. The RB is charging if $G_t-U_t \geq 0$, and discharging otherwise. 
The user's electricity consumption and the electricity price are assumed to be known for $H_F$ TSs beyond the current TS, naming $H_F$ as the \textit{prediction horizon}. Additionally, we assume that the EMU keeps memory about the past $H_P$ TSs, which we call the \textit{past horizon}. At each TS $t$, the EMU computes an EMP for the following $H_F$ TSs, using its knowledge of $U_t$ and the electricity cost within the prediction horizon, and its knowledge of the user load, grid load, and the RB level of energy within the past horizon.  

Instead of the LHM scenario, where full information about a consumer's future energy consumption is assumed to be known over the whole time period of interest, in this paper, we consider a more realistic scenario and assume that only partial knowledge of the consumer's future energy consumption is available to the EMU. The SHM assumption is motivated by the difficulty in obtaining reliable longer term predictions of a consumer's energy consumption.

\subsection{System Constraints}

Let $\overline{t+H_F} \triangleq \min\{t+H_F,N\}$. We do not allow wasting grid energy; that is, there are no battery overflows, i.e., we impose $B_{t-1} + (G_{t}-U_{t}) D \leq B_{\max}, \forall t$. This means that, at any time $t$ and considering a prediction horizon of $H_F$ TSs, the EMU has to satisfy the following constraint:
\begin{equation} \label{eq:batteryConstraint}
B_{t-1} + \sum_{s=t}^{\tau} (G_{s}-U_{s}) D \leq B_{\max}, 
\end{equation}
where $t \leq \tau \leq \overline{t+H_F}$.

While additional energy can be stored in the RB for future use, we do not allow demand rescheduling, so that user's energy demands are always satisfied at the time of request, i.e., we impose: $G_t D \geq U_t D - B_{t-1}$, $\forall t$. This leads to the following constraint for the EMU at time $t$:
\begin{equation} \label{eq:energySatisfied}
\sum_{s=t}^{\tau} (U_{s}-G_{s}) D \leq B_{t-1},
\end{equation}
where $t \leq \tau \leq \overline{t+H_F}$. This constraint is also implicitly verified by the equation expressing the evolution of the energy level in the battery:
\begin{equation}
0 \leq B_{t+1} = B_{t} + G_{t+1}D - U_{t+1}D.
\end{equation}

The power the RB can be charged or discharged at is constrained by $\hat{P}_{c}$ and $\hat{P}_{d}$, respectively. Thus, $\forall t$ we have:
\begin{align}
G_t - U_t &\leq \hat{P}_{c}, \label{eq:peakPowerCharging}\\
U_t  - G_t &\leq \hat{P}_{d}. \label{eq:peakPowerDischarging}
\end{align}

The model could be made more accurate by introducing further constraints, e.g., battery charging and discharging efficiency parameters, which we leave for future research focusing on the practical implications of the proposed UDS techniques.

We study and compare the two scenarios in which energy can or cannot be sold to the UP. The price of energy sold to the grid is set equal to the price of energy bought from it, i.e., we adopt the \emph{net metering} approach, in which the SM can measure bi-directional energy flows \cite{Payne:2000}. If energy cannot be sold, then
\begin{equation} \label{eq:ouputNonnegative}
G_t \geq 0, \qquad \forall t,
\end{equation}
whereas, if energy can be sold, we have:
\begin{equation} \label{eq:ouputNegative}
G_t: 
\begin{cases}
\geq 0,      &\text{if energy is purchased from the UP},\\
<0,   &\text{if energy is sold to the UP}.
\end{cases}
\end{equation}

Given $(U_t, B_t)=(u_t, b_t)$, $B_{\max}$ and the constraints (\ref{eq:peakPowerCharging})-(\ref{eq:ouputNonnegative}), the set of feasible energy requests at time $t$ is given by
\begin{multline}\label{eq:feasibleSetY}
\bar{\mathcal{G}_t}(u_t,b_t)  \triangleq \Big\{ g_t \in \mathcal{G}: 
\Big[u_t-\min\Big\{\frac{b_t}{D},\hat{P}_d\Big\}\Big]^+ \leq g_t  \\ \leq u_t + \min\Big\{\hat{P}_c,\frac{B_{\max}-b_t}{D}\Big\}\Big\}.
\end{multline}

If selling energy to the UP is allowed, then the feasible set is as in Eq. (\ref{eq:feasibleSetY}), but without the $[\cdot]^+$ operator.

The EMP computes the grid load at each TS while satisfying the above constraints. We consider a model predictive control approach, whereby the user load and the cost of energy are known beforehand within the prediction horizon $[t+1,\ldots,t+H_F]$, and the goal is to jointly minimize the information leaked about a user's energy consumption as well as the cost the user incurs to purchase energy from the UP. While non-causal knowledge of the electricity price for the typical range of interest is a realistic assumption in today's energy networks, non-causal knowledge of power consumption is appropriate for appliances whose activity can be accurately predicted, e.g., refrigerators, boilers, heaters and electric vehicles. We note that the setting studied in \cite{Giaconi:2017SGC}, which assumes all future energy consumption and cost information to be known beforehand, is a lower bound on the setting studied in this paper, as more information leads to a better privacy-cost trade-off.

Let the target load at time $t$ be denoted by $W_t$. We measure the privacy leakage as the average variance of the grid load $G^N$ from the target load profile $W^N$
\begin{equation}\label{eq:privacy}
\mathcal{P} \triangleq \frac{1}{N}\sum_{t=1}^N (G_t - W_t)^2,
\end{equation}
according to which, perfect privacy is achieved when $G_t = W_t$, $\forall t$. We adopt squared distance in (\ref{eq:privacy}) not to discriminate between negative and positive deviations of $G_t$ from $W_t$. The average cost incurred by the user is given by
\begin{equation}\label{eq:cost}
\mathcal{C} \triangleq \frac{1}{N} \sum_{t=1}^N G_t C_t,
\end{equation}
where $C_t$ is the cost of energy at time $t$, which is determined by the specific ToU tariff employed by the UP.

\subsection{Simulation Settings}

We use real SM consumption traces from the UK Dale dataset \cite{UKDALE}. We convert the original resolution of $6$ to $10$ minutes to reduce the computational complexity. We consider a Tesla Powerwall $2$ \cite{tesla} as RB, for which $B_{\max}=13.5$kWh, and $\hat{P}_c=\hat{P}_d=5$kW. We consider a ToU tariff that was offered in the UK \cite{tide_new}, in which the off-peak price is $4.99$p/kWh during 23:00 to 6:00, the medium price is $11.99$p/kWh during 6:00 to 16:00 and during 19:00 to 23:00, and the peak price is $24.99$p/kWh during 16:00 to 19:00. All the simulation results are obtained for a time interval spanning 14 consecutive days, to average over a considerably large amount of data. For simplicity, we will mostly present numerical results when selling energy to the grid is not allowed, unless energy selling leads to significantly different results.


\section{Target Load as a Constant Value}\label{sec:constantTargetLoad}

In this section, following up on \cite{Kalogridis:2010SGC,Yang:2015TSG} and \cite{Tan:2017TIFS}, we assume that the goal of the EMU is to keep the grid load as constant as possible. In \cite{Giaconi:2017TIFS} it is assumed that all the future user load and energy cost values are known, and the EMU can fix a target value for the whole duration, e.g., one whole day. In our model the information available to the EMU on $U_t$ and $C_t$ is limited to the prediction horizon, and changes over time; thus, the target load cannot be constant, and its variability depends on the length of the past and prediction horizons. In this section, given the knowledge of the cost of energy and the user's power consumption, the aim is to characterize both the optimal target load $W^*$ and the optimal grid load $G^*$ so as to optimize the overall privacy-cost trade-off.

\begin{figure}[!t]
\begin{subfigure}[t]{.48 \columnwidth}
\centering
\includegraphics[width=1\columnwidth]{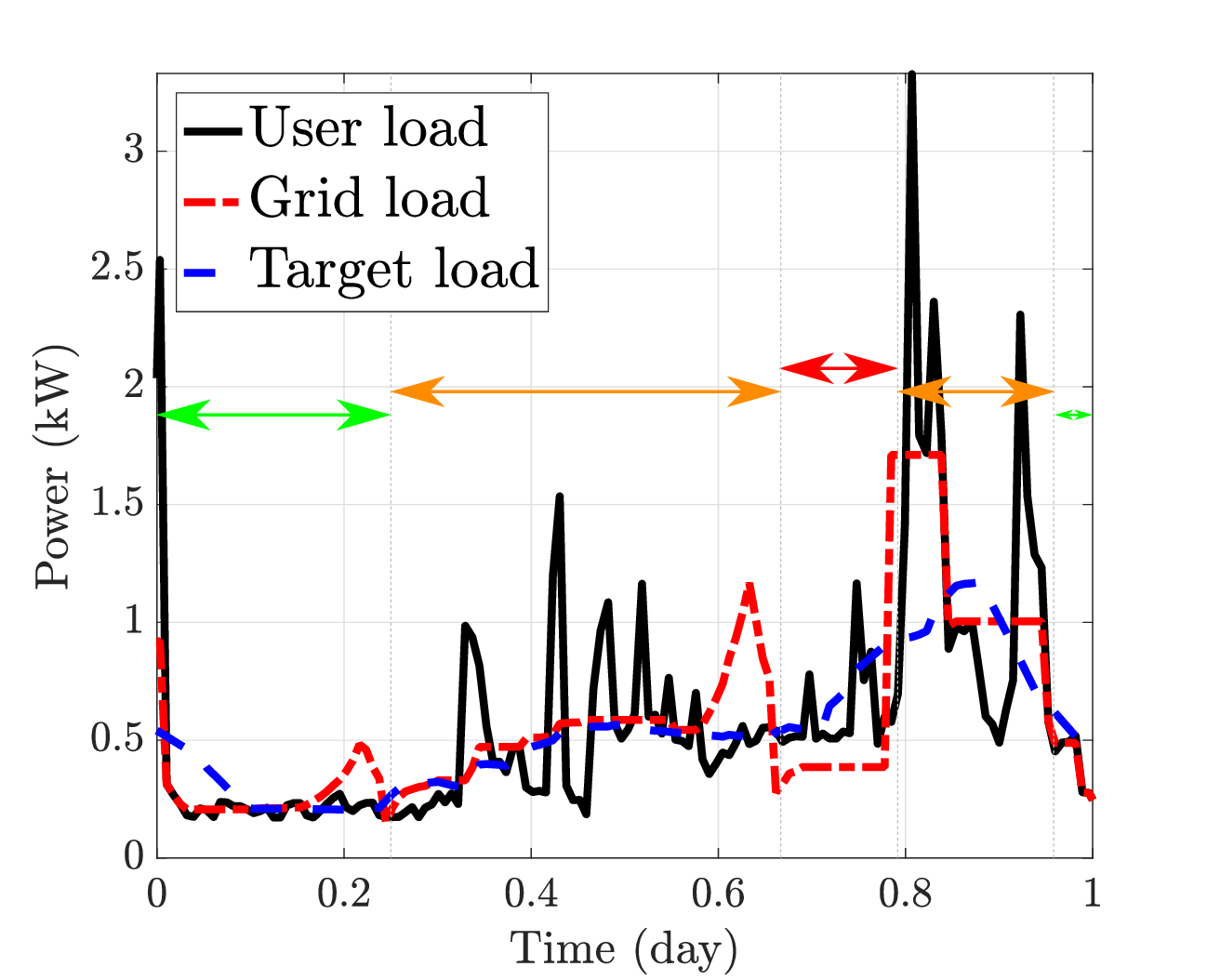}
\caption{SHM, no energy selling.}
\label{fig:constantShort}
\end{subfigure}
\begin{subfigure}[t]{.48\columnwidth}
\centering
\includegraphics[width=1\columnwidth]{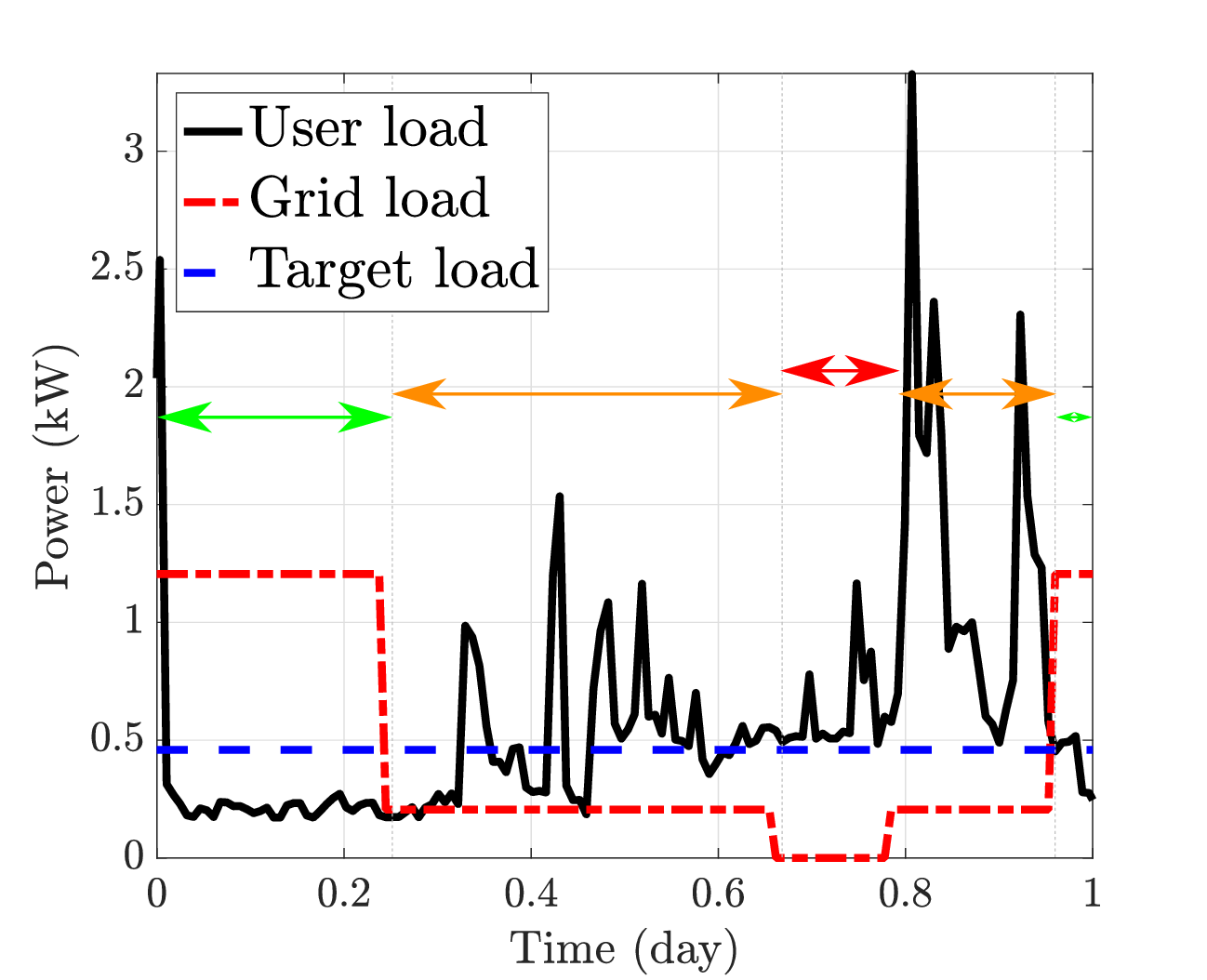}
\caption{LHM, no energy selling.}
\label{fig:constantLong}
\end{subfigure} \\
\begin{subfigure}[t]{.48\columnwidth}
\centering
\includegraphics[width=1\columnwidth]{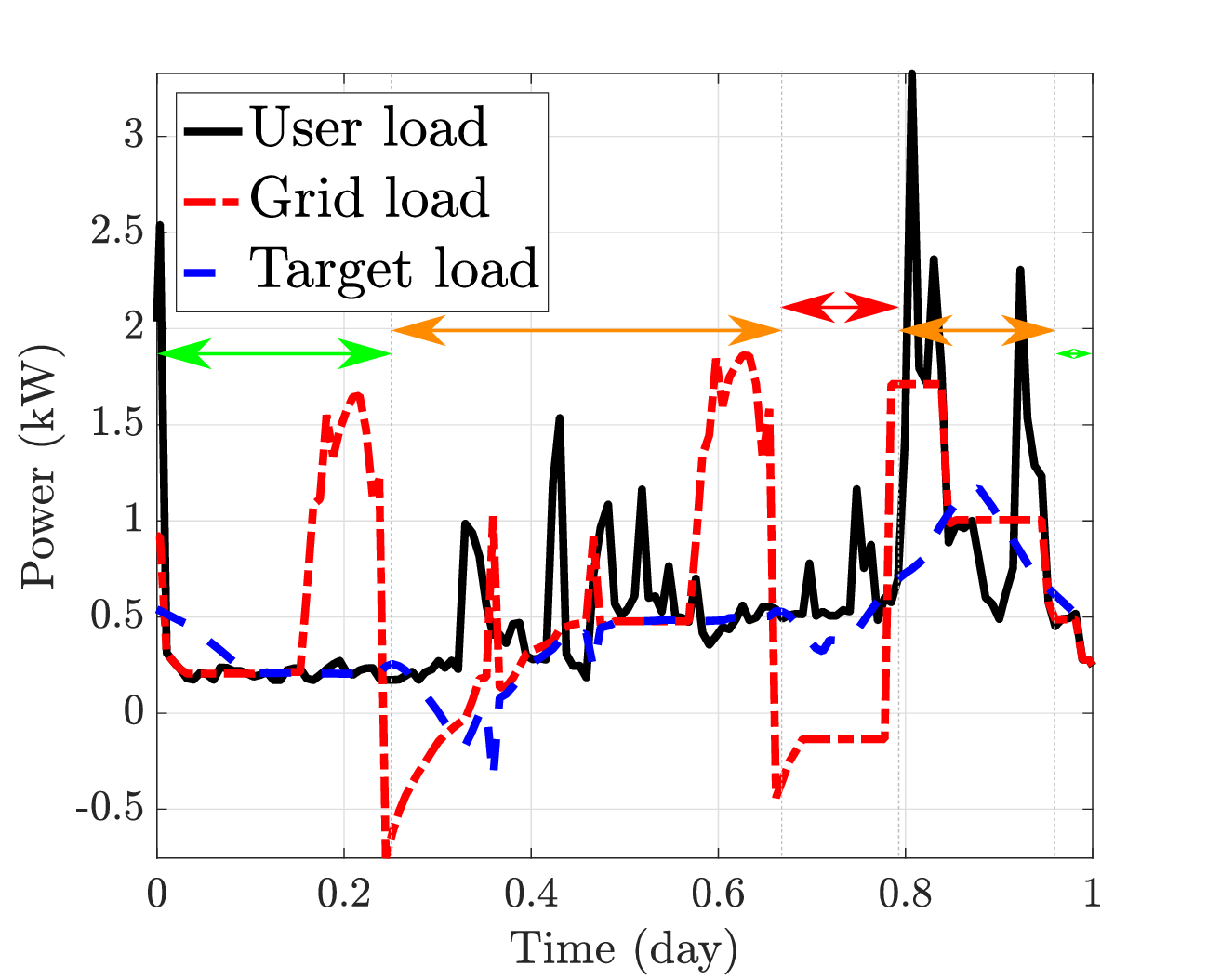}
\caption{SHM, energy selling.}
\label{fig:constantShortS}
\end{subfigure}
\begin{subfigure}[t]{.48\columnwidth}
\centering
\includegraphics[width=1\columnwidth]{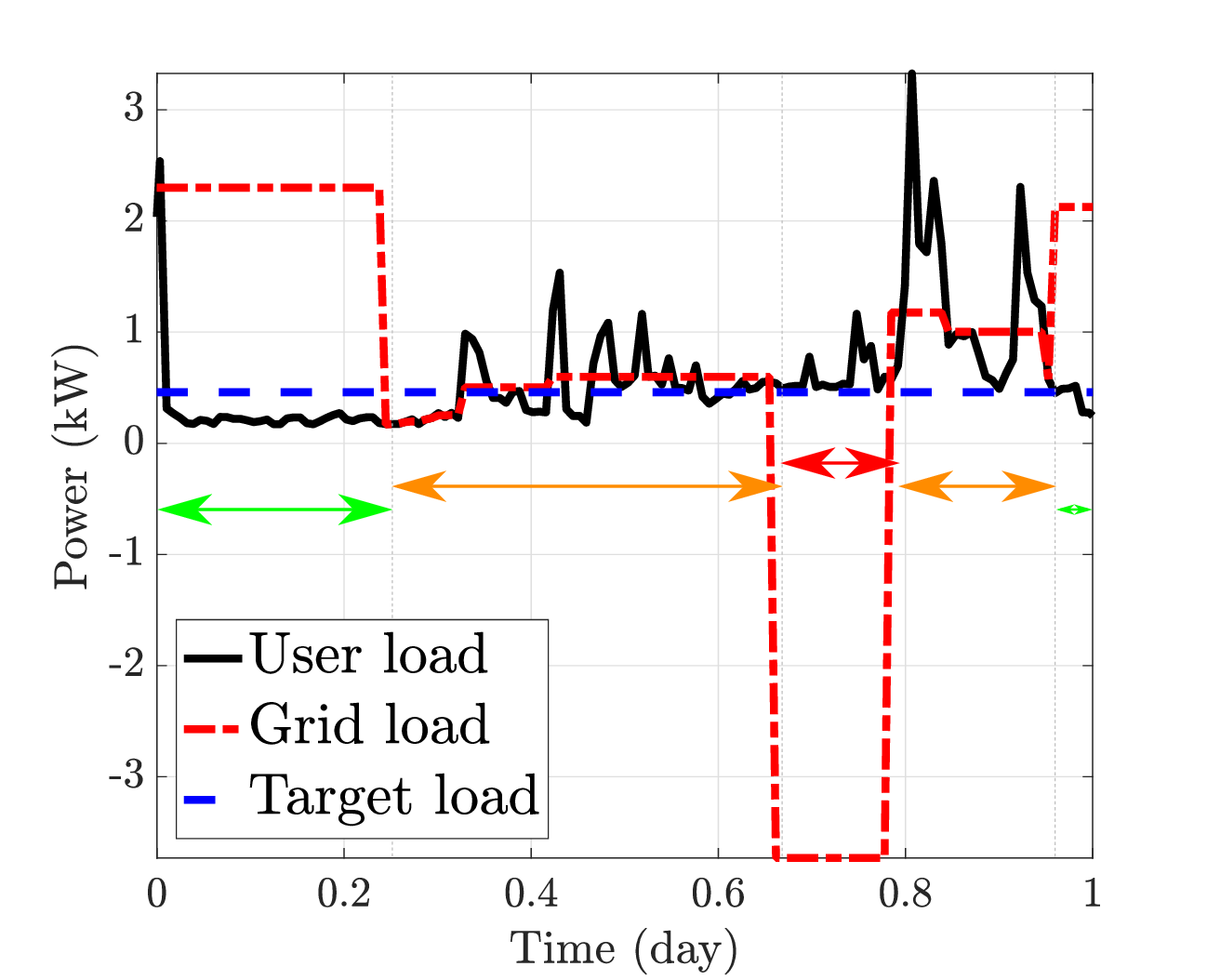}
\caption{LHM, energy selling.}
\label{fig:constantLongS}
\end{subfigure}
\caption{Power profiles for $\alpha=0.5$ and $H_F=H_P=2$h. In the figures, the arrows of green, orange and red colors denote time intervals characterized by off-peak, medium and peak price for the electricity cost, respectively.}\label{fig:constantLongVSShort}
\end{figure}

\begin{figure*}[!b]
\setcounter{equation}{11}
\hrulefill
\begin{multline}\label{eq:lagrangian}
  \mathcal{L}(G_t^{\overline{t+H_F}},W_t,\bm{\lambda}) =  \alpha  \sum_{\tau=\overline{t-H_P}}^{\overline{t+H_F}} (G_{\tau} - W_t)^2 + (1-\alpha)\sum_{\tau=t}^{\overline{t+H_F}} G_{\tau} C_{\tau} + \sum_{\tau=t}^{\overline{t+H_F}} \lambda_{\tau}^{(1)} \bigg[ D \sum_{s=t}^{\tau} (G_s-U_s) - B_{\max}  + B_{t-1}\bigg]\\
 +  \sum_{\tau=t}^{\overline{t+H_F}} \lambda_{\tau}^{(2)} \bigg[ D  \sum_{s=t}^{\tau} (U_s-G_s) - B_{t-1}\bigg]   +  \sum_{\tau=t}^{\overline{t+H_F}} \lambda_{\tau}^{(3)} (G_\tau - U_\tau - \hat{P}_c) + \sum_{\tau=t}^{\overline{t+H_F}} \lambda_{\tau}^{(4)} (U_\tau -G_\tau - \hat{P}_d).
\end{multline}
\setcounter{equation}{10}
\end{figure*}

Given the nature of the objective functions and the constraints, pairs of ($\mathcal{P}$, $\mathcal{C}$) form a convex region and the optimal points can be characterized by the Pareto boundary of this region. Hence, the objective can be cast as a weighted sum of privacy leakage (\ref{eq:privacy}) and cost (\ref{eq:cost}):
\begin{equation}\label{eq:ConstantShort}
\min_{G_t^{\overline{t+H_F}},W_t} \alpha \sum_{\tau=\overline{t-H_P}}^{\overline{t+H_F}} (G_{\tau} - W_t)^2 + (1-\alpha)\sum_{\tau=t}^{\overline{t+H_F}} G_{\tau} C_{\tau},
\end{equation}
where $0 \leq \alpha \leq 1$ is the weighting parameter, i.e., if $\alpha=0$ only cost of energy is minimized, whereas if $\alpha=1$ only information leakage is minimized; and $\overline{t-H_P} \triangleq \max\{t-H_P,0\}$. We remark that setting the value of $\alpha$ is up to the consumer, who is in charge of deciding whether to focus more on protecting her privacy or on saving costs. The result of the minimization in Eq. (\ref{eq:ConstantShort}) is the grid load for the current TS and the entire duration of the prediction horizon $G_t^{\overline{t+H_F}}$, and the target load $W_t$. Eq.  (\ref{eq:ConstantShort}) characterizes the target load value $W_t$ for the finite prediction horizon, which leads to the optimal privacy-cost trade-off over this horizon, based on the available information. At TS $t+1$, the minimization (\ref{eq:ConstantShort}) is carried out again based on the additional information that becomes available, i.e., $G_{t+H_F+1}$ and $C_{t+H_F+1}$, and $G_{t+1}$ and $W_{t+1}$ are determined. The past horizon  $\sum_{\tau=\overline{t-H_P}}^{t-1} (G_{\tau} - W_{t})^2$ is considered when optimizing for the privacy objective, since it ensures smoother variations of the overall target load profile. We note that, since privacy and cost in Eq. (\ref{eq:ConstantShort}) may have significantly different magnitudes, they need to be further normalized to get the Pareto optimal solution consistent with $\alpha$ \cite{Narzisi:2008}. 

\begin{figure}[!ht]
\begin{subfigure}[t]{.48\columnwidth}
\centering
\includegraphics[width=1\columnwidth]{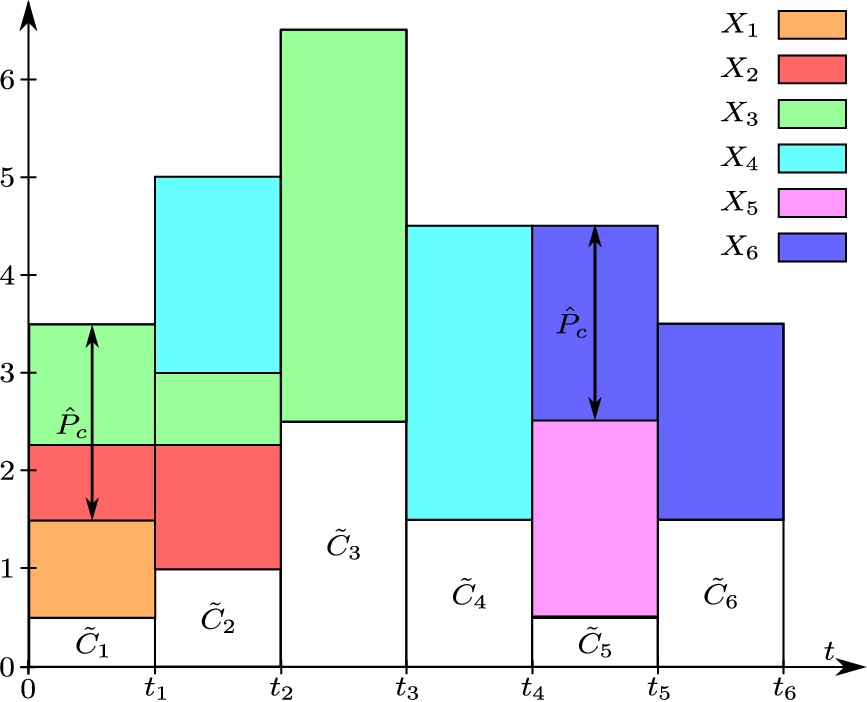}
\caption{$\alpha=0$.}
\label{fig:water0}
\end{subfigure}
\begin{subfigure}[t]{.48\columnwidth}
\centering
\includegraphics[width=1\columnwidth]{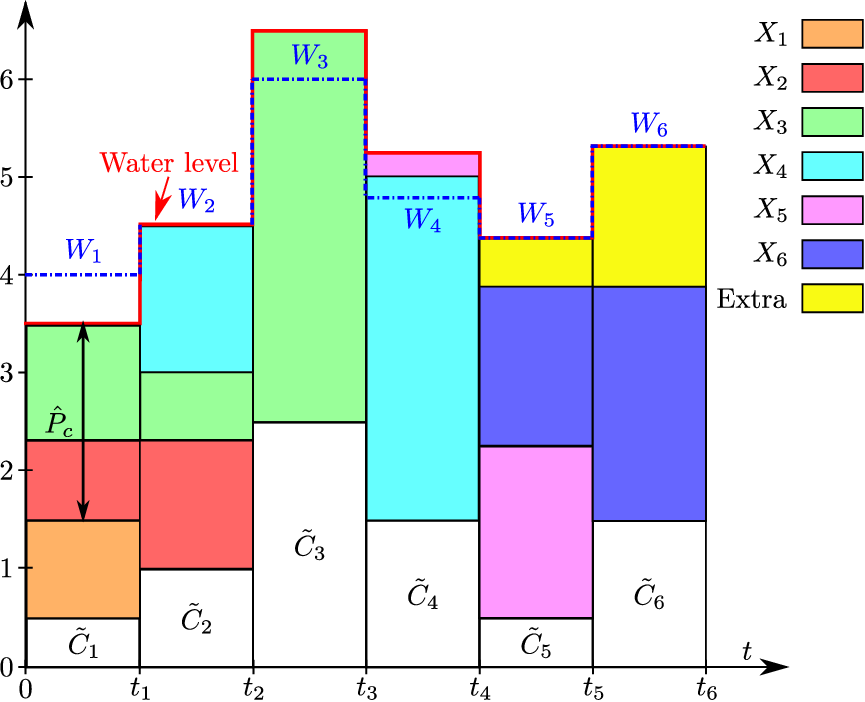}
\caption{$\alpha=1$.}
\label{fig:water1}
\end{subfigure}
\caption{Optimal grid and target load profiles, for $B_{\max}=4$, $\hat{P}_c=\hat{P}_d=2$, when energy selling is not allowed.}
\label{fig:PVconstant_water}
\end{figure}

\begin{remark}
Differently from \cite{Giaconi:2017SGC}, here we do not impose the RB to be emptied at the end of each time window $[t-H_P,t+H_F]$, since here the end of the prediction horizon does not typically coincide with the end of the time horizon of interest, and the energy remaining in the RB can be utilized in the following TSs. Since the algorithm jointly minimizes privacy leakage and cost, the RB is normally emptied at the end of the time horizon of interest $N$, unless $\alpha$ is high. If $\alpha \rightarrow 1$ and the RB is large, a sustained demand of energy may take place in short term, which is ultimately constrained by $B_{\max}$ and $\hat{P}_c$.
\end{remark}

Fig. \ref{fig:constantLongVSShort} shows the load profiles of SHM and LHM over one day. As expected, the LHM provides better performance as the resultant profile is much flatter and hides most of the consumption spikes, which typically reveal more information about user's behavior. However, also the SHM leads to a reasonable suppression of consumption peaks, despite relying on much less data. Also, the SHM reveals more information about the low-frequency variation of user's energy consumption, which, however, is common across households, and thus provides only a limited amount of personal information. Moreover, Figs. \ref{fig:constantShort} and \ref{fig:constantShortS} show that the peaks of the grid load generated by SHM are not necessarily aligned with those of the user load.

\subsection{Solution to the Optimization Problem (\ref{eq:ConstantShort})}

In the following we consider the optimization problem (\ref{eq:ConstantShort}) for $\alpha\neq\{0,1\}$. We analyze first the optimal solution when selling of energy is allowed, i.e., (\ref{eq:ouputNonnegative}) does not hold. Based on (\ref{eq:ConstantShort}) and the constraints (\ref{eq:batteryConstraint})-(\ref{eq:energySatisfied}) and (\ref{eq:peakPowerCharging})-(\ref{eq:peakPowerDischarging}), we define the Lagrangian function in (\ref{eq:lagrangian}) at the bottom of the next page, where $\lambda_{\tau}^{(j)} \geq 0$, for $1\leq j\leq 4$, are the Lagrange multipliers, and $t\leq \tau \leq \overline{t+H_F}$. Denoting the vectors in bold, we have  $\bm{\lambda}=[\bm{\lambda}^{(1)},\bm{\lambda}^{(2)},\bm{\lambda}^{(3),},\bm{\lambda}^{(4)}]$. The slackness conditions are imposed on the inequality constraints, for $\tau=t,t+1,\ldots,\overline{t+H_F}$:
\setcounter{equation}{12}
\begin{align}
  &\lambda_{\tau}^{(1)} \bigg[ D \sum_{s=t}^{\tau} (G_s-u_s) - B_{\max} + B_{t-1}\bigg]=0, \label{eq:slack1}\\
  &\lambda_{\tau}^{(2)} \bigg[ D \sum_{s=t}^{\tau} (U_s-G_s) - B_{t-1}\bigg] = 0, \label{eq:slack2} \\
  &\lambda_{\tau}^{(3)} (G_\tau - U_\tau - \hat{P}_c)=0, \label{eq:slack3}\\
  &\lambda_{\tau}^{(4)} (U_\tau -G_\tau - \hat{P}_d)=0. \label{eq:slack4} 
\end{align}

Let  $a_{\tau} \triangleq D \sum_{s=\tau}^{\overline{t+H_F}}(\lambda_{s}^{(2)}-\lambda_{s}^{(1)}) - \lambda_\tau^{(3)}+\lambda_\tau^{(4)}$, and $\tilde{C}_{\tau} \triangleq \frac{(1-\alpha)}{2\alpha}C_{\tau}$. Applying the Karush-Kuhn-Tucker (KKT) conditions and setting the gradient of the Lagrangian to zero, we obtain the following expressions:
\begin{align}
G^{*}_{\tau} &= \frac{a_{\tau}}{2\alpha} - \tilde{C}_{\tau} + W_t^*, \qquad \text{for } \tau=t,\ldots,\overline{t+H_F},  \label{eq:optimalY} \\ 
W_t^{*} &= \frac{\sum_{\tau=\overline{t-H_P}}^{\overline{t+H_F}} G_\tau^*}{1+\min\{H_P,t\}+\min\{H_F,N-t\}}. \label{eq:optimalWi}
\end{align}

\begin{figure}[!t]
\begin{subfigure}[t]{.48\columnwidth}
\centering
\includegraphics[width=1\columnwidth]{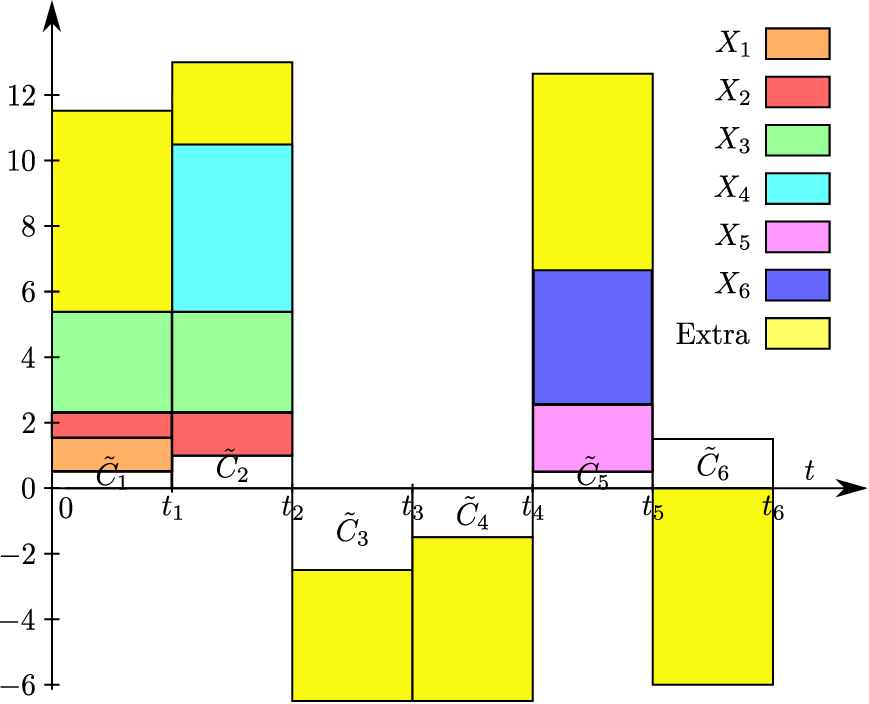}
\caption{$\alpha=0$.}
\label{fig:waterS}
\end{subfigure}\hfill
\begin{subfigure}[t]{.48\columnwidth}
\centering
\includegraphics[width=1\columnwidth]{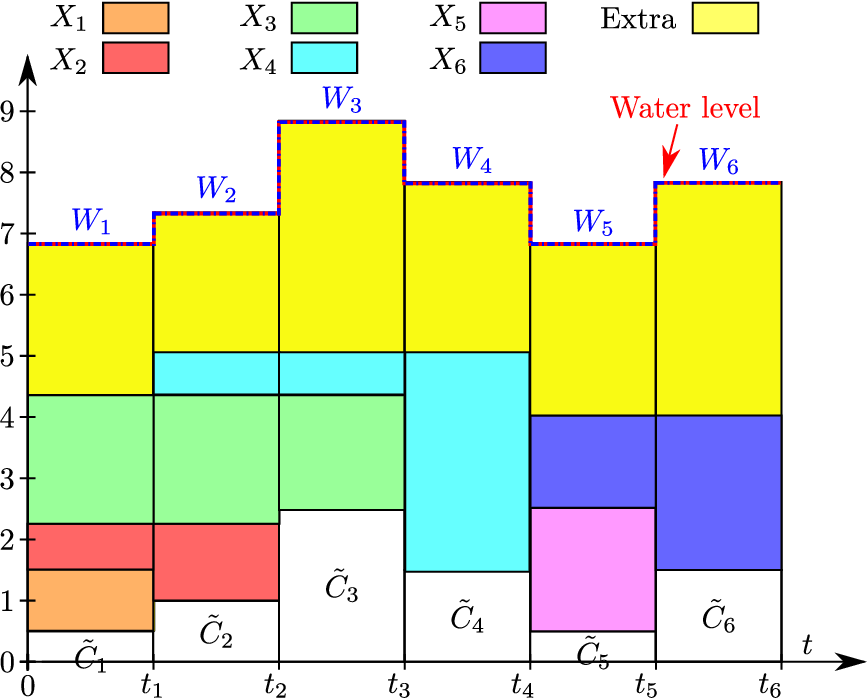}
\caption{$\alpha=1$.}
\label{fig:waterS1}
\end{subfigure}
\caption{Optimal grid and target load profiles, for $B_{\max}=20$, $\hat{P}_c=\hat{P}_d=10$, when energy selling is allowed.}
\label{fig:PVconstant_water_selling}
\end{figure}

\begin{figure*}[t!]
\begin{subfigure}[c]{.33\textwidth}
\centering
\includegraphics[width=1\columnwidth]{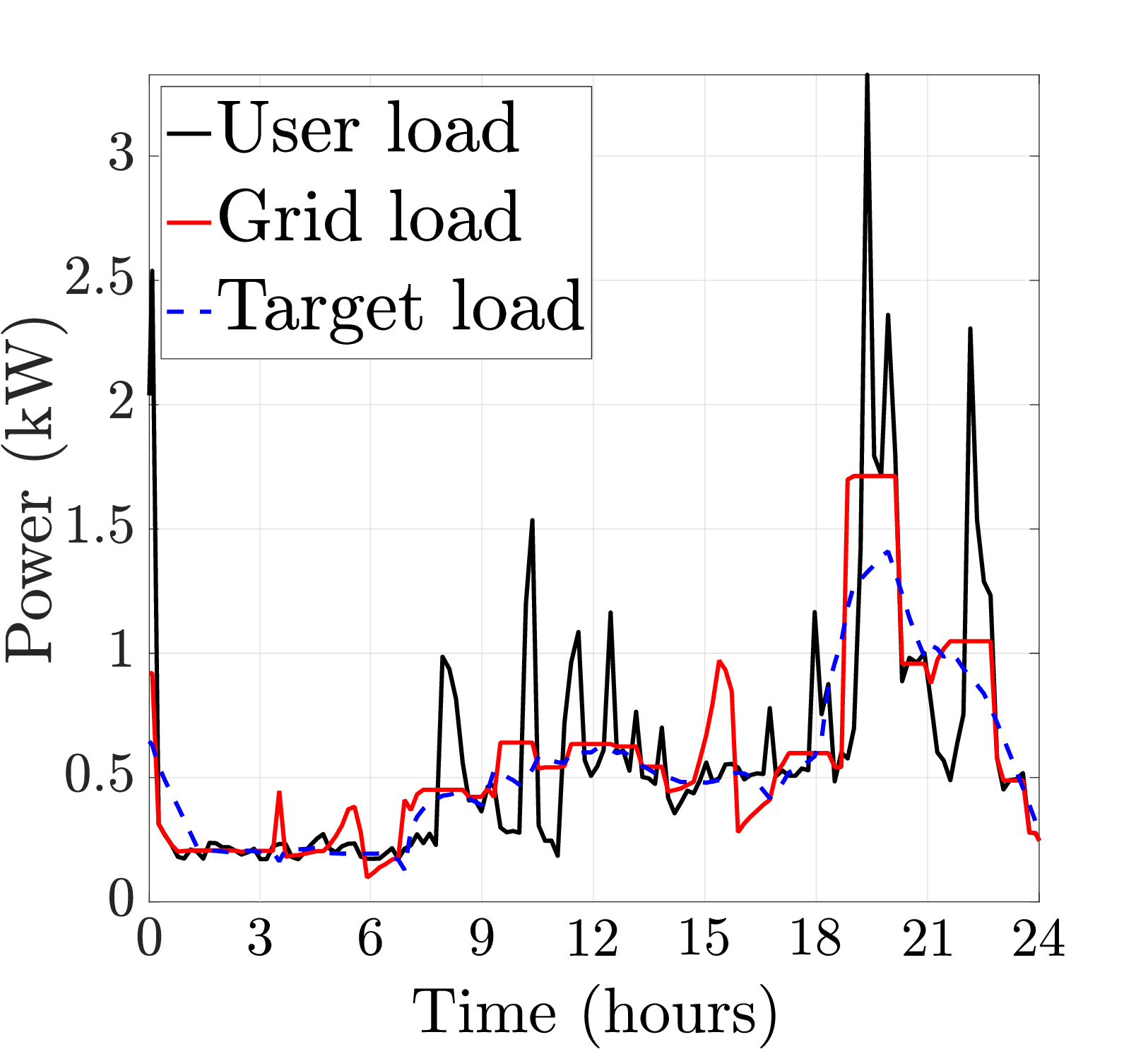}
\caption{$H_P=H_F=1$h.}
\label{fig:PVconstant_Past1}
\end{subfigure}
\begin{subfigure}[c]{.33\textwidth}
\centering
\includegraphics[width=1\columnwidth]{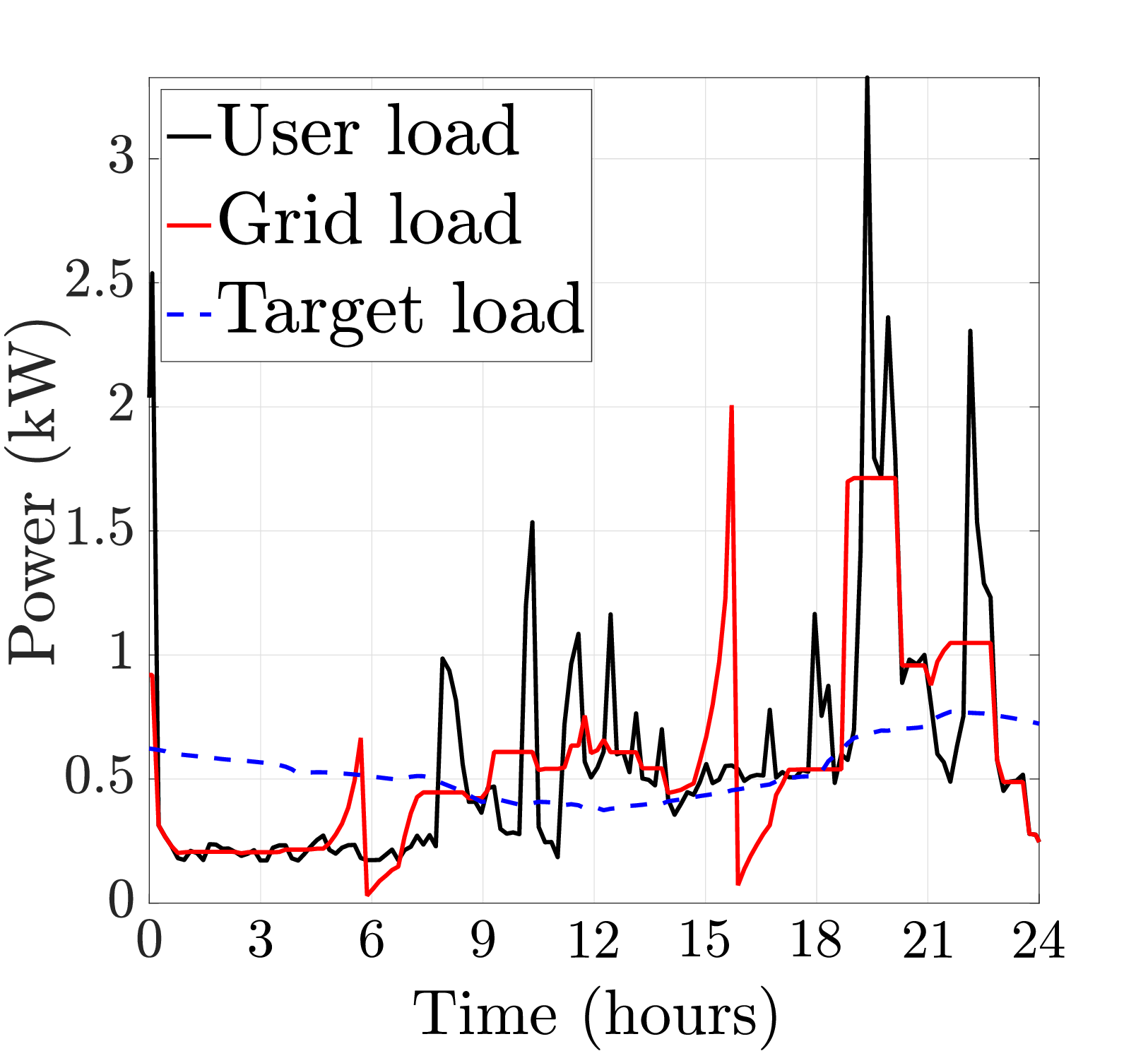}
\caption{$H_P=12$h, $H_F=1$h.}
\label{fig:PVconstant_Past2}
\end{subfigure}
\begin{subfigure}[c]{.33\textwidth}
\includegraphics[width=1\columnwidth]{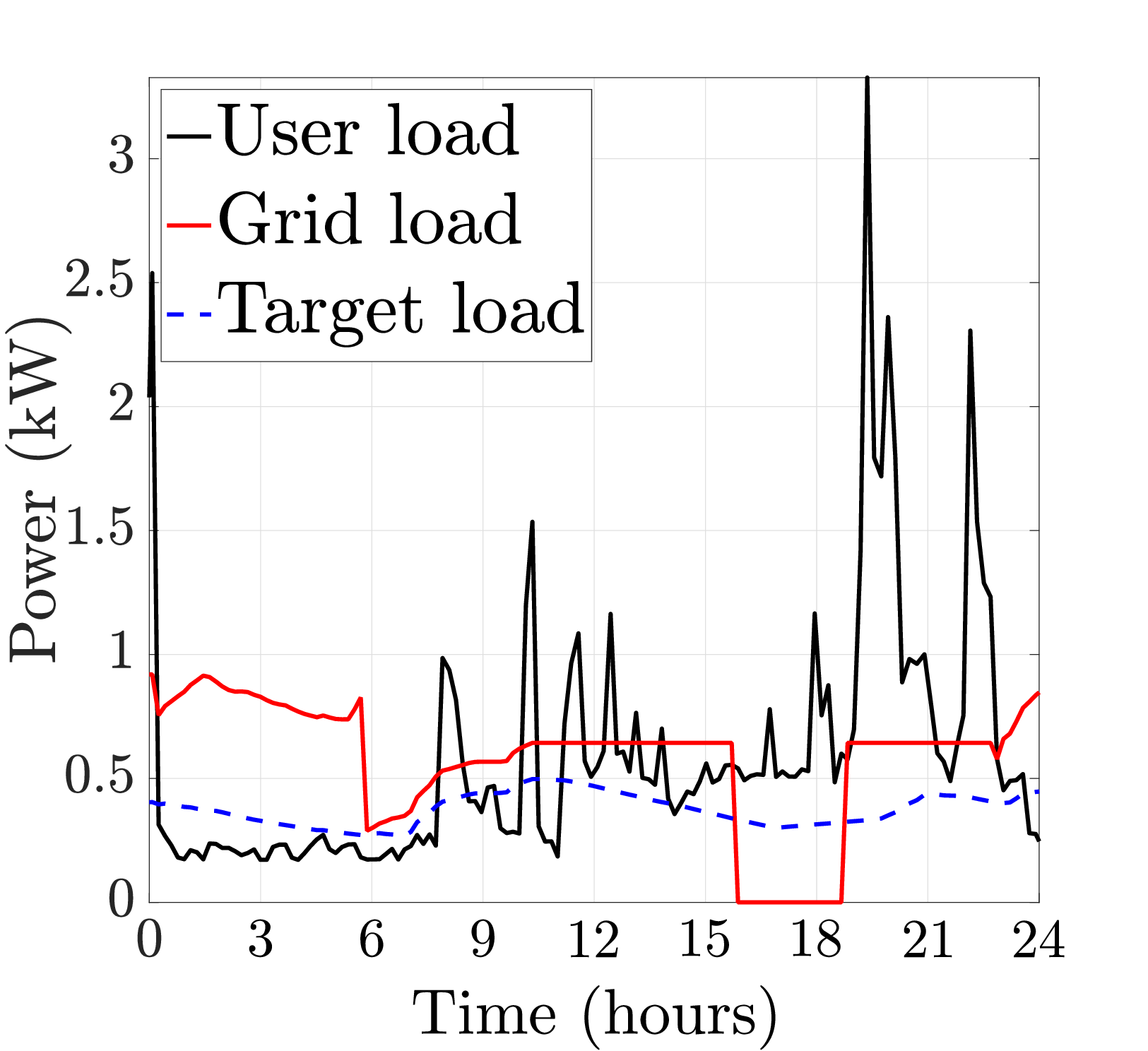}
\caption{$H_P=1$h, $H_F=12$h.}
\label{fig:PVconstant_Pred}
\end{subfigure}
\caption{Comparison between various past and prediction horizons for $\alpha=0.5$, when energy selling is not allowed.}
\label{fig:PVconstant_PastShortCompare}
\end{figure*}

The optimal solution for the grid load given in (\ref{eq:optimalY}) resembles the classical water-filling algorithm \cite{Cover:1991}. However, differently from the classical water-filling formulation, here the water level, $G^{*}_\tau+ \tilde{C}_t =\frac{a_\tau}{2\alpha} + W_t^{*}$, is not constant, but varies over time due to the instantaneous power constraints. The optimal solutions given by Eqs. (\ref{eq:optimalY}) and (\ref{eq:optimalWi}) depend on the values of the Lagrangian multipliers and can be determined numerically. 

When $\alpha=0$, the only objective is to minimize the cost, and Eq. (\ref{eq:ConstantShort}) reduces to a linear program, which can be solved using standard linear programming solvers. On the other hand, when the user is not concerned about the cost, i.e., $\alpha=1$, Eq. (\ref{eq:ConstantShort}) leads to a quadratic program analogous to the general case.

When energy selling is not allowed, the constraint (\ref{eq:ouputNonnegative}) holds, and the Lagrangian in (\ref{eq:lagrangian}) is modified accordingly.
The slackness conditions are given in Eqs. (\ref{eq:slack1})-(\ref{eq:slack4}), as well as, for $\tau=t,\ldots,\overline{t+H_F}$: $\lambda_{\tau}^{(5)} G_\tau=0$, and $\lambda^{(6)} W_t=0$.
Let $\tilde{a}_\tau \triangleq D \sum_{s=\tau}^{\overline{t+H_F}}(\lambda_{s}^{(2)}-\lambda_{s}^{(1)}) - \lambda_\tau^{(3)}+\lambda_\tau^{(4)}+\lambda_\tau^{(5)}$. Then, we obtain the following expressions, counterparts of (\ref{eq:optimalY}) and (\ref{eq:optimalWi}):
\begin{align}
G^{*}_\tau &= \bigg[\frac{\tilde{a}_\tau}{2\alpha} - \tilde{C}_\tau + W_t^*\bigg]^+, \label{eq:optimalYNoSel}\quad \text{for } \tau=t,\ldots,\overline{t+H_F}, \\ 
W_t^{*}&= \Bigg[\frac{\sum_{\tau=\overline{t-H_P}}^{\overline{t+H_F}} G_\tau^* + \lambda^{(6)}}{1+\min\{H_P,t\}+\min\{H_F,N-t\}}\Bigg]^+. \label{eq:optimalWiNoSel}
\end{align}

\subsection{Illustration of the Water-filling Solution}

Here we present the solution for some simple scenarios to acquire an intuition on the solution of the optimization problem (\ref{eq:ConstantShort}), and on its water-filling interpretation. Assume energy selling is not allowed, $N=6$, and $D=1$, i.e., power and energy can be used interchangeably. Consider $u^6=[1,2,6,5,2,4]$ and $c^6=[1,2,5,3,1,3]$, $H_P=H_F=2$ TSs, and an RB with $B_{\max}=4$ and $\hat{P}_{c}=\hat{P}_{d}=2$. Fig. \ref{fig:water0}  shows the optimal solution for $\alpha=0$, $G^{*,6}=[3,4,4,3,4,2]$. Since the electricity is cheaper for $t=\{1,2\}$, more energy is requested from the grid at these TSs and stored in the RB to satisfy the demand at later TSs. However, such energy is limited by $B_{\max}$ and $\hat{P}_{c}$. At $t=1$, $\hat{P}_{c}$ limits the grid load, since $G^*_1=u_1+\hat{P}_{c}=3$, and the level of energy in the RB at the end of the first TS is $B_1=\hat{P}_{c}=2$. At $t=2$, the grid load is limited by $\hat{P}_c$ and $B_{\max}$ simultaneously, and $G^*_2=u_2+\hat{P}_{c}=4$, and $B_2=B_{\max}=4$. Note that, although the third TS is the most expensive, $G^*_3=4$ because the RB cannot be discharged by more than $2$ units of energy ($\hat{P}_d=2$). For the same reason, $G^*_4=3$, while the remaining energy was stored in the second TS. Similar considerations hold for the last two TSs. Fig. \ref{fig:water1} illustrates the optimal solution for $\alpha=1$, $G^{*,6}=[3,3.5,4,3.75,3.88,3.81]$. In this scenario, the EMP tries to match $G_t$ to the target load $W_t$, even at the cost of asking more energy than needed. The energy demand at $t=1$ is the same as the case $\alpha=0$, whereas at $t=2$ less energy is stored in the battery to be used at $t=4$, so that the water level matches the target load in this TS (at the expense of a higher cost at $t=4$). It is noteworthy that more energy than needed is requested at $t=4$, and used to satisfy the demand during at $t=5$, which is cheaper. Finally, more energy than needed is requested at $t=\{4,5\}$ (depicted in yellow), allowing the EMP to match the target and the grid load. Such energy is stored in the RB for future use.

\begin{figure}[!t]
\begin{subfigure}[t]{.33\columnwidth}
\centering
\includegraphics[width=1\columnwidth]{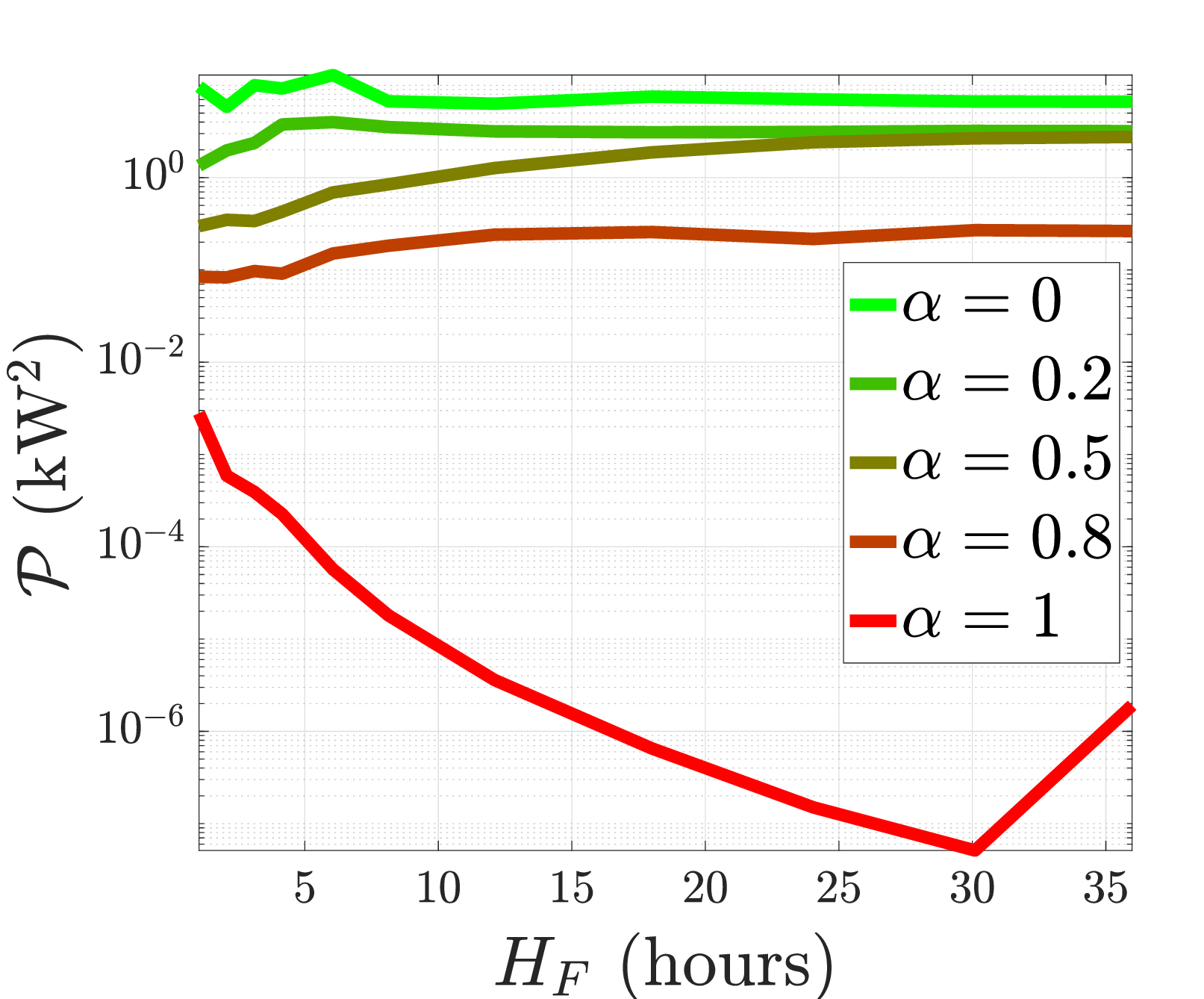}
\caption{Energy selling.}
\label{fig:constant_VaryingPred_Leakage}
\end{subfigure}\hfill
\begin{subfigure}[t]{.33\columnwidth}
\centering
\includegraphics[width=1\columnwidth]{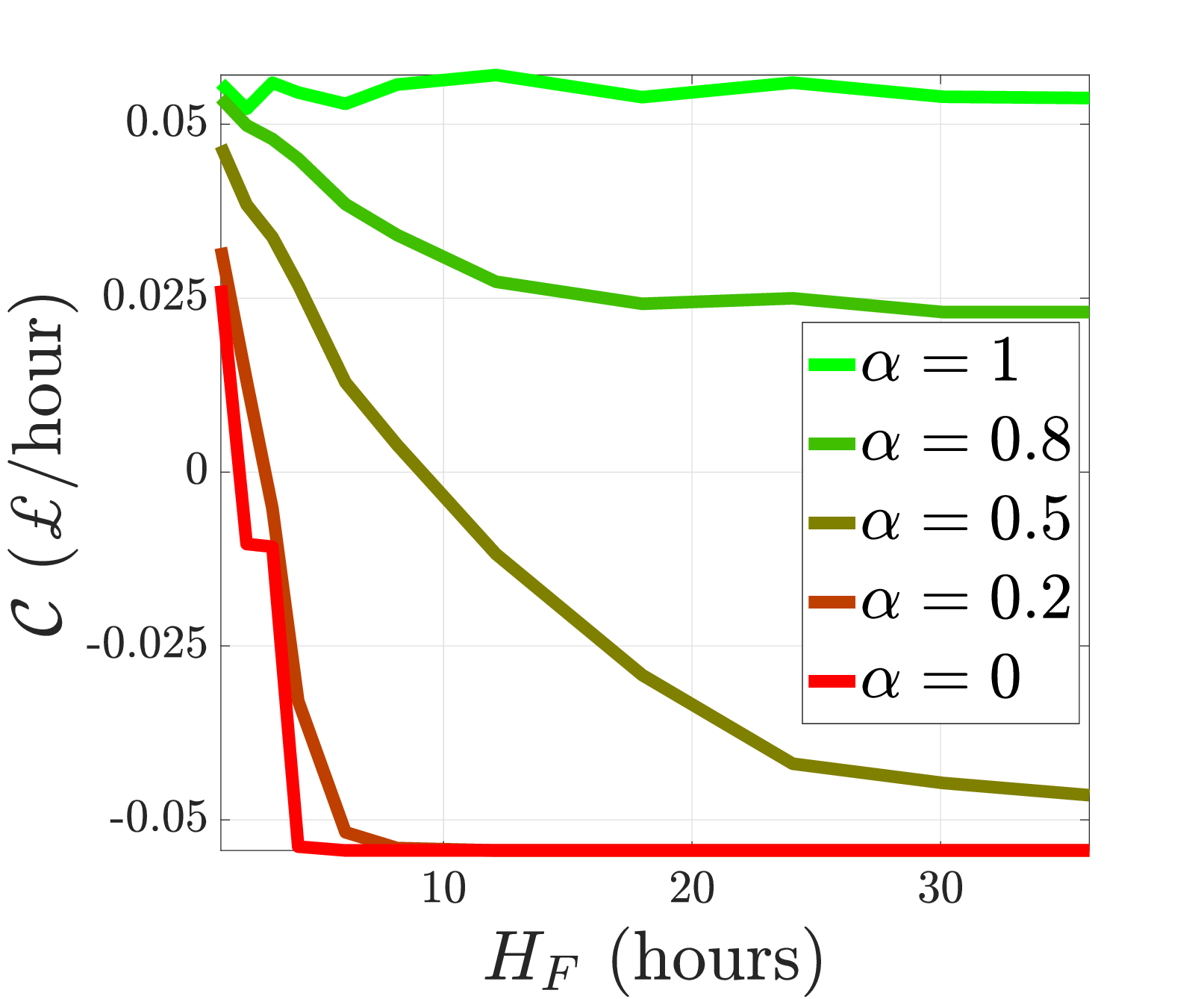}
\caption{Energy selling.}
\label{fig:constant_VaryingPred_Cost}
\end{subfigure}\hfill
\begin{subfigure}[t]{.33\columnwidth}
\centering
\includegraphics[width=1\columnwidth]{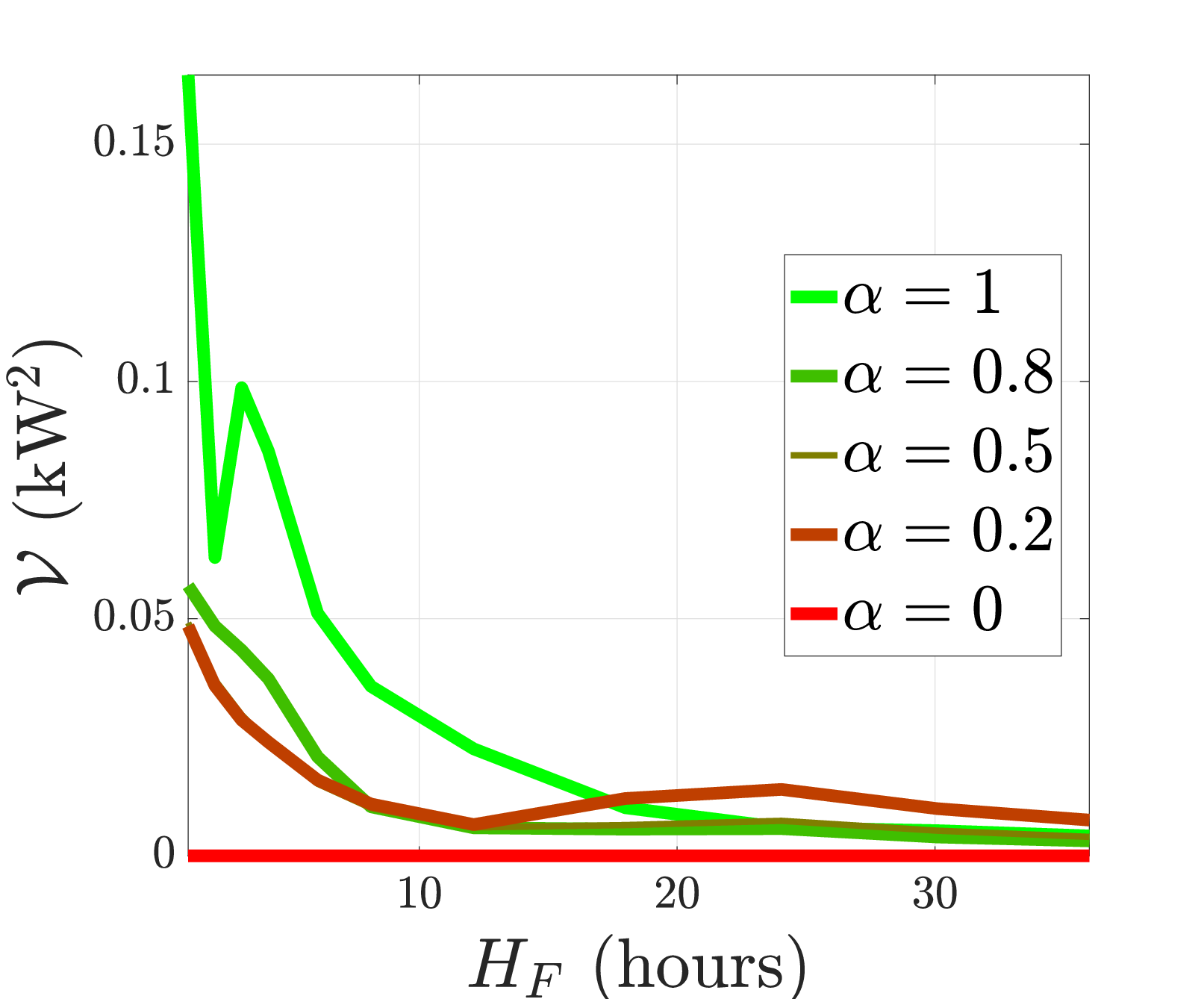}
\caption{No energy selling.}
\label{fig:constant_VaryingPred_targetVariance}
\end{subfigure}
\caption{Impact of the prediction horizon $H_F$ on leakage, cost and target load variance ($H_P=2$h).}
\label{fig:PVconstant_VaryingPred_CostCompare}
\end{figure}

\begin{figure}[!t]
\begin{subfigure}[t]{.33\columnwidth}
\centering
\includegraphics[width=1\columnwidth]{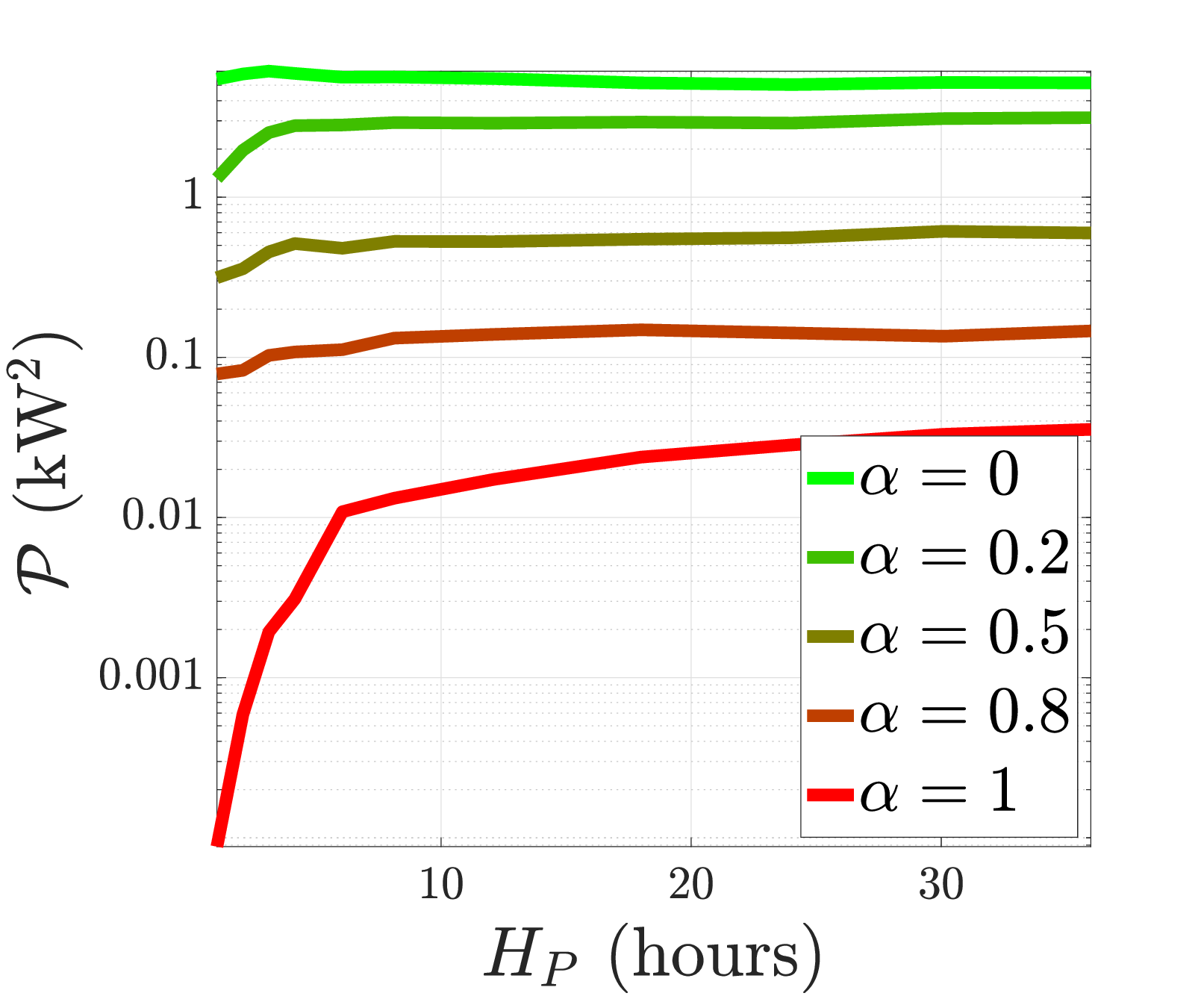}
\caption{Energy selling.}
\label{fig:PVconstant_VaryingPast_Leakage}
\end{subfigure}\hfill 
\begin{subfigure}[t]{.33\columnwidth}
\centering
\includegraphics[width=1\columnwidth]{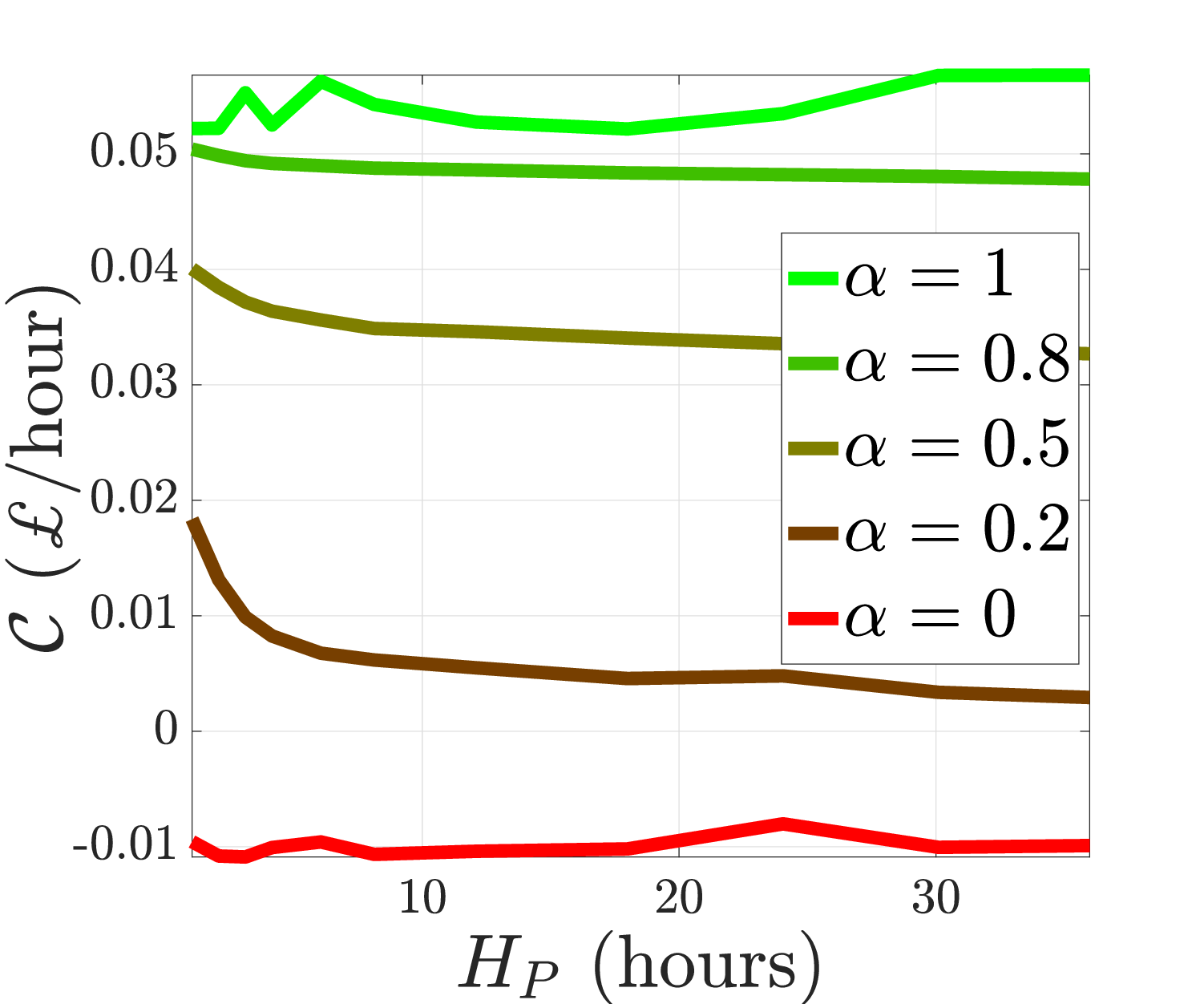}
\caption{Energy selling.}
\label{fig:PVconstant_VaryingPast_Cost}
\end{subfigure}\hfill
\begin{subfigure}[t]{.33\columnwidth}
\centering
\includegraphics[width=1\columnwidth]{Figures/summaryTargetVariancePred_noselling_rescaled.eps}
\caption{No energy selling.}
\label{fig:PVconstant_VaryingPast_TargetVarianceCompare}
\end{subfigure}
\caption{Impact of the past horizon $H_P$ on leakage, cost and target load variance ($H_F=2$h).}
\label{fig:PVconstant_VaryingPast_VarianceCompare}
\end{figure}

\begin{figure}[!t]
\begin{subfigure}[t]{.5\columnwidth}
\centering
\includegraphics[width=1\columnwidth]{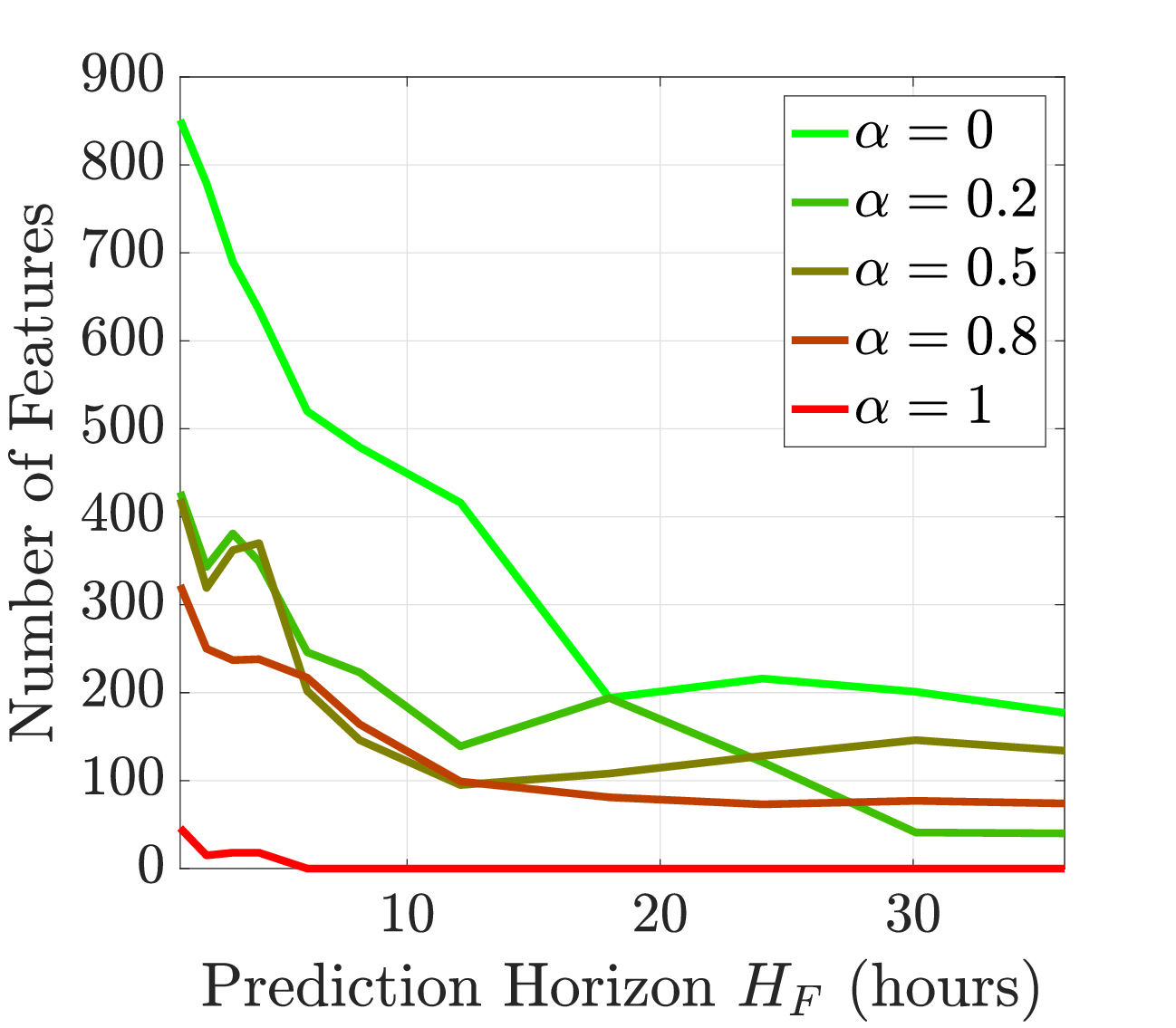}
\caption{$H_P=2$h.}
\label{fig:constant_VaryingPred_features}
\end{subfigure}\hfill
\begin{subfigure}[t]{.5\columnwidth}
\centering
\includegraphics[width=1\columnwidth]{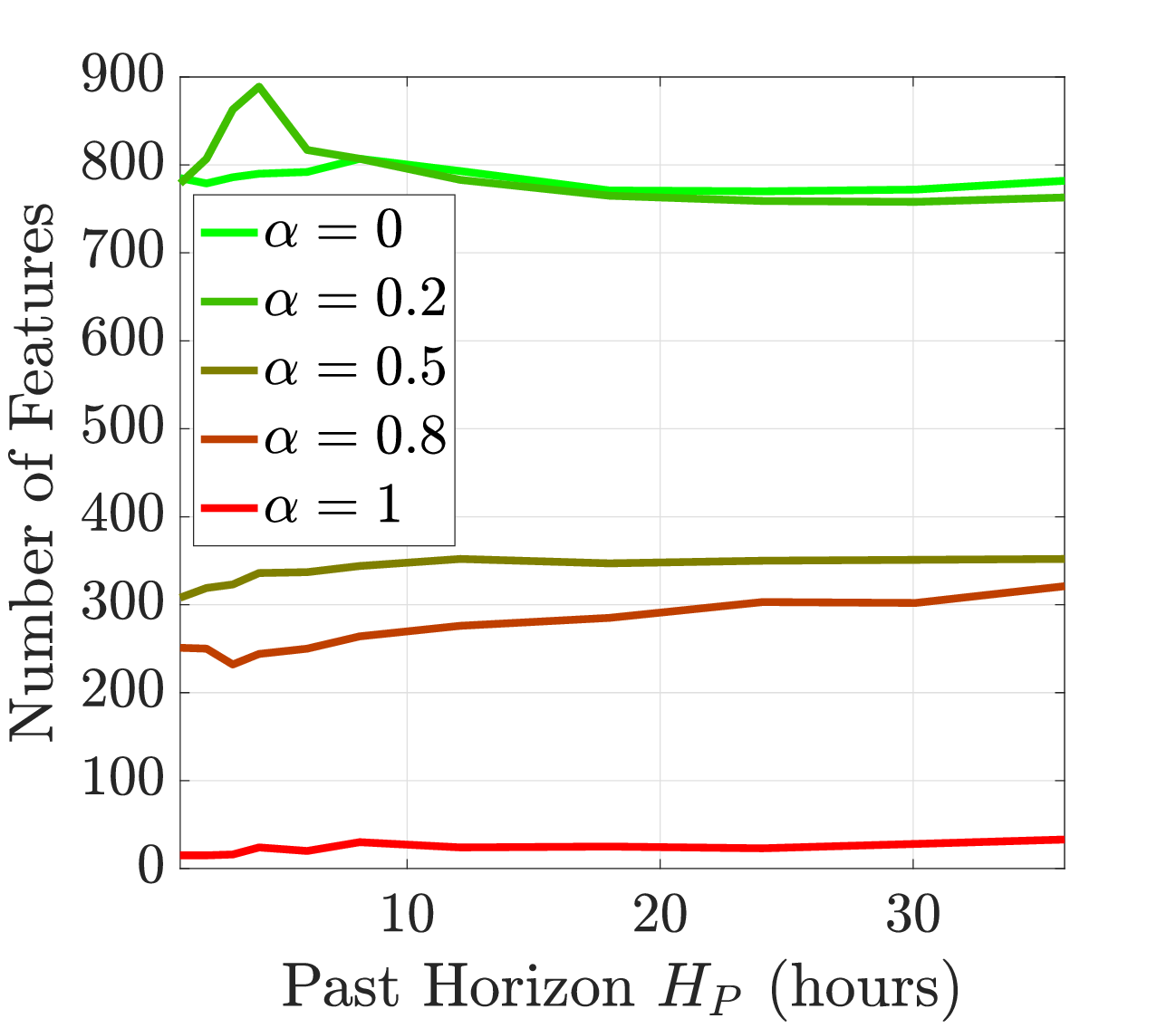}
\caption{$H_F=2$h.}
\label{fig:constant_VaryingPast_features}
\end{subfigure}
\caption{Number of features vs $H_F$ and $H_P$, no energy selling.}
\label{fig:constant_VaryingSummary}
\end{figure}

\begin{figure*}[!t]
\begin{subfigure}[t]{.33\textwidth}
\includegraphics[width=1\columnwidth]{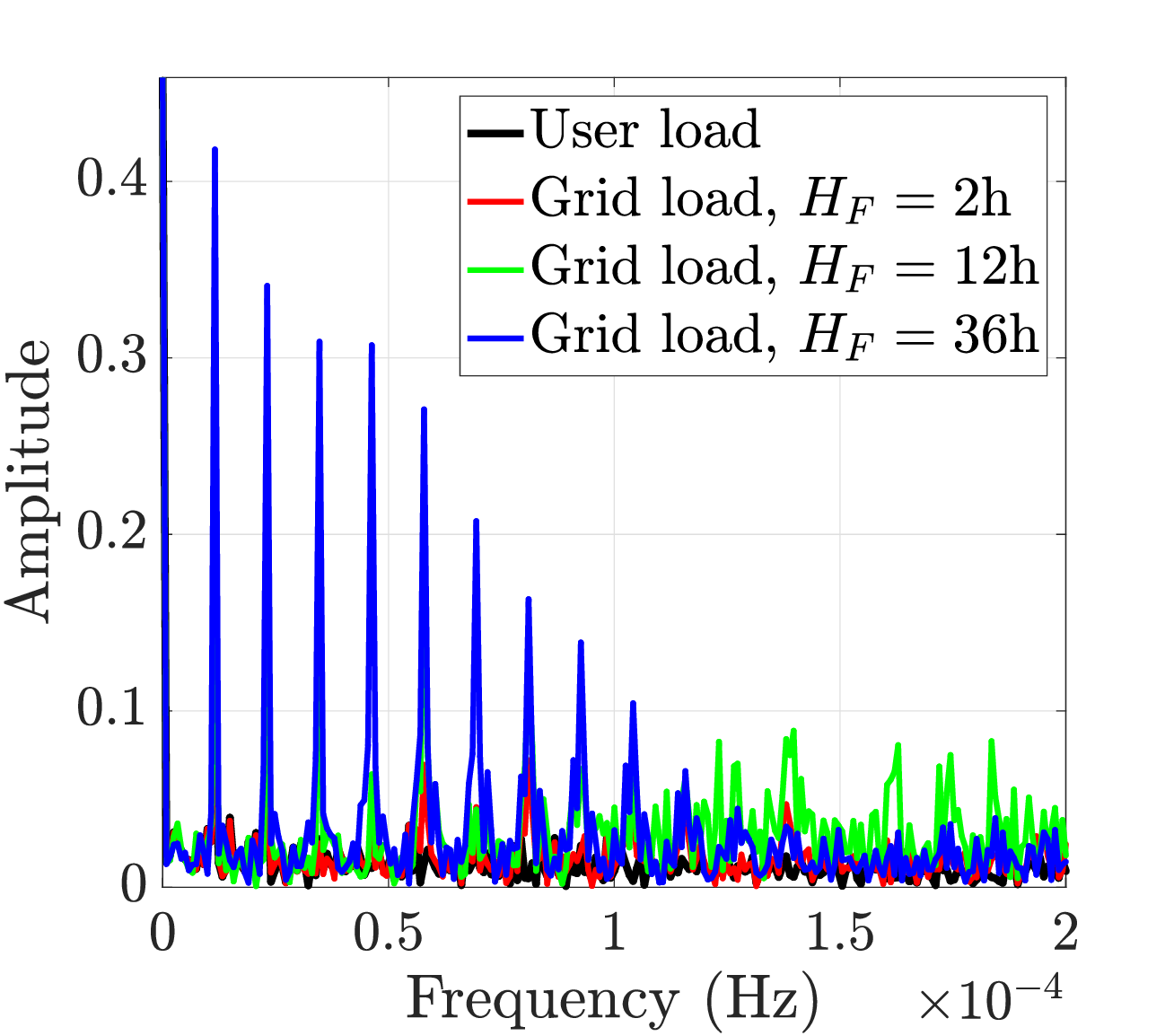}
\caption{$\alpha=0$, $H_P=2$h.}
\label{fig:PVconstant_VaryingPred_Spectra0selling}
\end{subfigure}\hfill
\begin{subfigure}[t]{.33\textwidth}
\centering
\includegraphics[width=1\columnwidth]{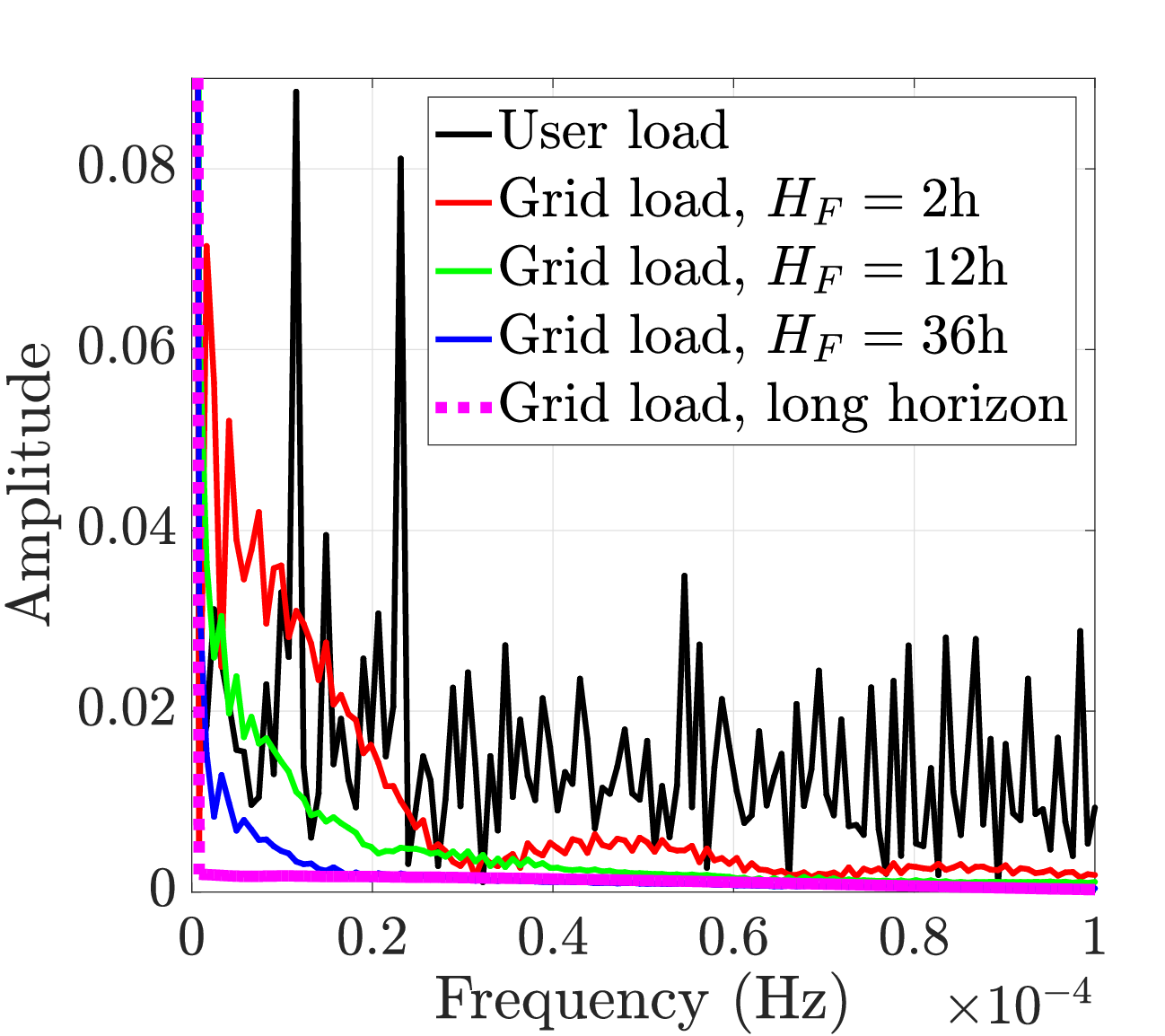}
\caption{$\alpha=1$, $H_P=2$h.}
\label{fig:PVconstant_VaryingPred_Spectra1}
\end{subfigure}\hfill
\begin{subfigure}[t]{.33\textwidth}
\centering
\includegraphics[width=1\columnwidth]{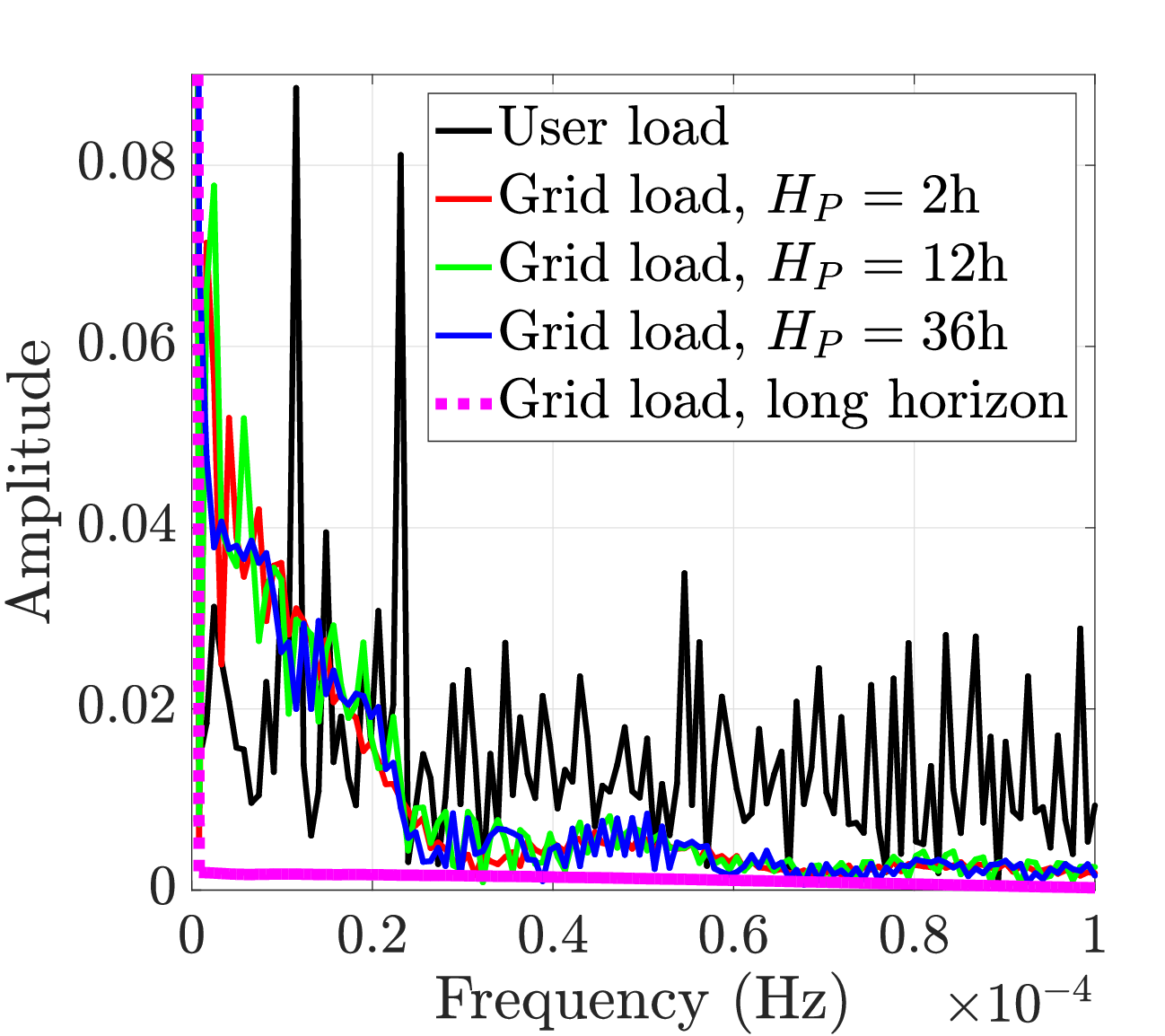}
\caption{$\alpha=1$, $H_F=2$h.}
\label{fig:PVconstant_VaryingPast_Spectra1_Sellingc}
\end{subfigure}
\caption{Power spectra vs $H_F$ and $H_P$, and no energy selling.}
\label{fig:constant_Spectra}
\end{figure*}

Fig. \ref{fig:PVconstant_water_selling} shows the same scenario with $B_{\max}=20$, $\hat{P}_c=\hat{P}_d=10$, and when energy can be sold. When $\alpha=0$, Fig. \ref{fig:waterS} shows that energy is bought when it is cheaper and sold back to the grid when it is more expensive, maximising user's profit. $\tilde{C}_3$ and $\tilde{C}_4$ are plotted as negative since energy is sold in these TSs. In fact, the RB is emptied of the energy stored during the first two TSs at the end of the fourth TS, and it is emptied again of the energy stored during the fifth TS at the end of the sixth TS. When $\alpha=1$, Fig. \ref{fig:waterS1} shows that a larger RB permits greater flexibility but also boosts the amount of energy requested. This is not necessarily a disadvantage, as such energy can be used at a later TS.

\subsection{Impact of the Duration of Prediction and Past Horizons}

Fig. \ref{fig:PVconstant_PastShortCompare} shows the load profiles for various combinations of $H_F$ and $H_P$. As expected, a larger $H_F$ produces flatter target and grid loads (see Fig. \ref{fig:PVconstant_Pred}), compared to a smaller $H_F$ (see Fig. \ref{fig:PVconstant_Past1}). However, when $H_F$ is larger the resulting grid load is more distant from the target load, thus resulting in a higher information leakage according to our definition of privacy leakage in Eq. (\ref{eq:privacy}). In fact, when $H_F$ is small, the grid load values that are compared to the target load are few, and the EMP is able to determine a target load that is close to the grid load within the analyzed time window. On the contrary, when $H_F$ is large, the EMP needs to find a single target load that matches a longer interval of grid load values; as a result, the target load may be less representative for some periods. A larger $H_P$ leads to a flatter target load (see Fig. \ref{fig:PVconstant_Past2}), however, the grid load is spikier compared to considering a larger $H_F$.

\begin{figure}[!t]
\begin{subfigure}[c]{.5\columnwidth}
\centering
\includegraphics[width=1\columnwidth]{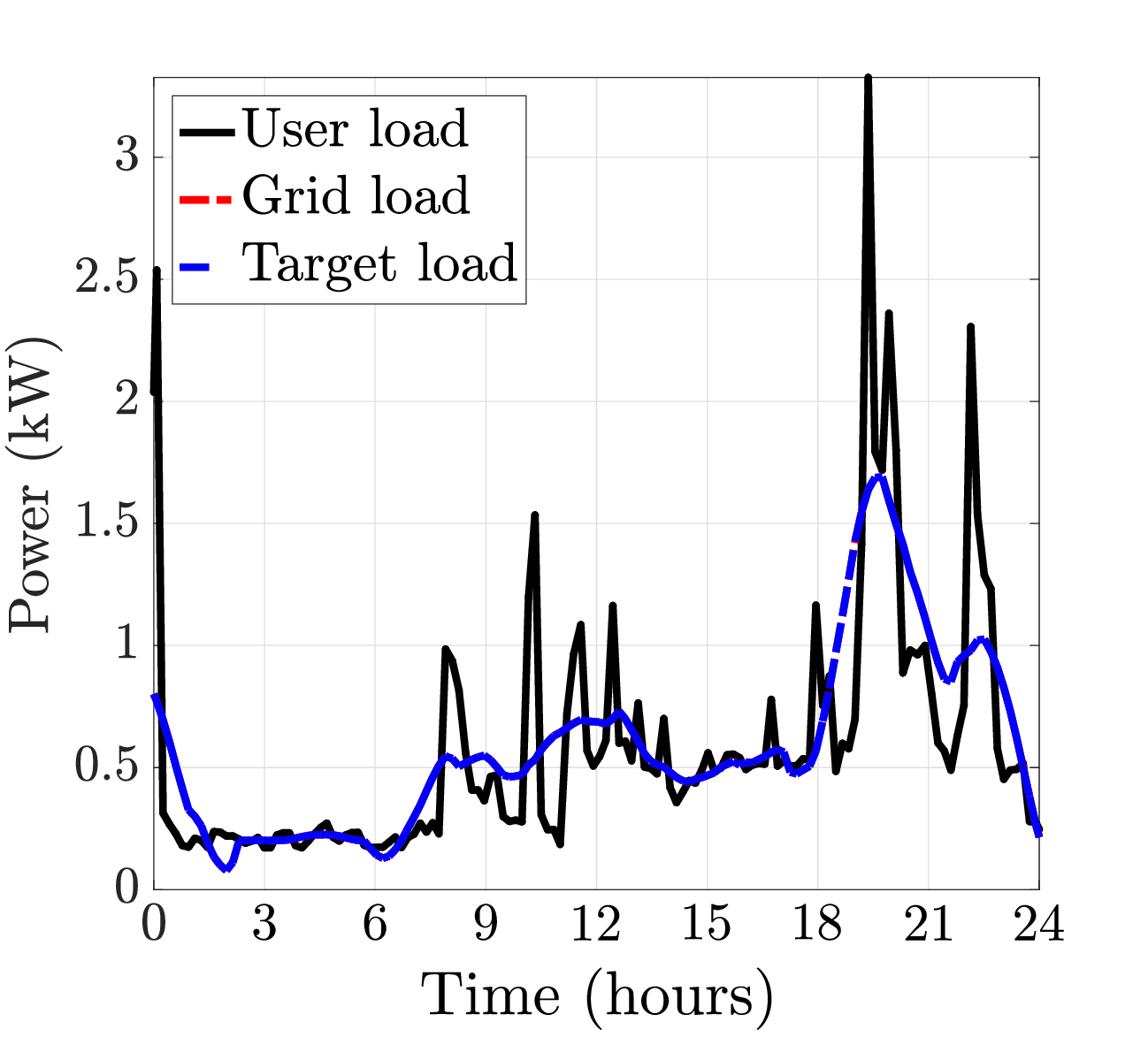}
\caption{SHM.}
\label{fig:filterShort}
\end{subfigure}\hfill
\begin{subfigure}[c]{.5\columnwidth}
\includegraphics[width=1\columnwidth]{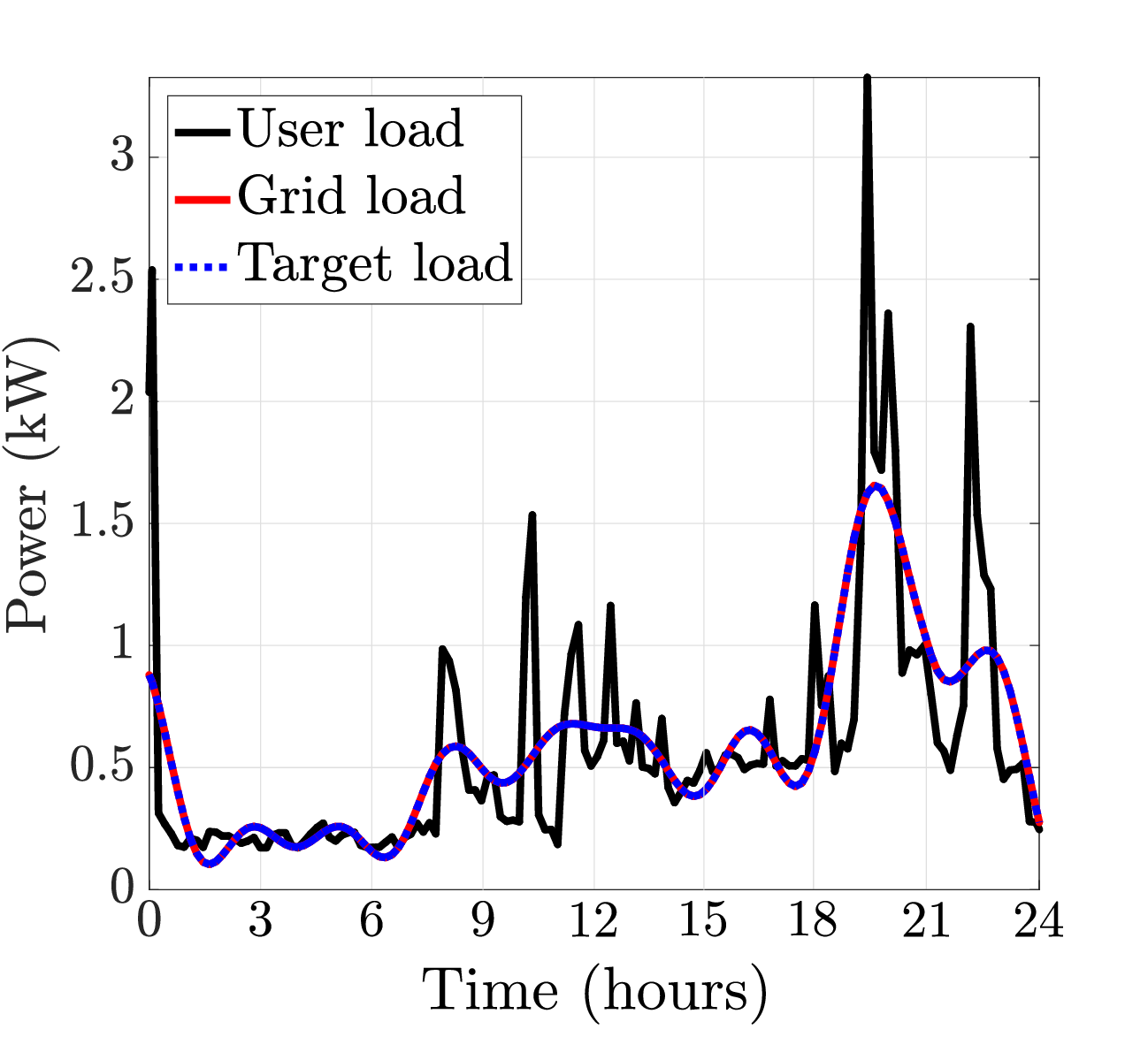}
\caption{LHM.}
\label{fig:filterLong1}
\end{subfigure}
\caption{Filtered target load scenario with cut-off frequency of $0.1$mHz, $\alpha=1$, and $H_F=H_P=2$h.}
\label{fig:filterLongVSShort_Filter}
\end{figure}

Fig. \ref{fig:PVconstant_VaryingPred_CostCompare} shows the average information leakage $\mathcal{P}$, average cost $\mathcal{C}$, and average target load variance with respect to $H_F$. The average target load variance, which can be considered as another privacy indicator, is defined as 
\begin{equation}
\mathcal{V}\triangleq\frac{1}{N}\sum_{t=1}^{N}(W_t-\mu_W)^2,
\end{equation}
where $\mu_W$ is the mean of $W$ over time. When $\alpha$ is small, i.e., the main focus is to minimize the cost, a larger $H_F$ reduces the average cost up to a certain extent, beyond which it cannot be further reduced (see Fig. \ref{fig:constant_VaryingPred_Cost}); whereas for $\alpha=1$ the cost does not change considerably with $H_F$. Opposite considerations hold for the information leakage, which even slightly increases when $\alpha\neq1$ (see Fig. \ref{fig:constant_VaryingPred_Leakage}). This is due to the fact that a longer prediction horizon generates a grid load that is more distant from the target load, except for $\alpha=1$, when the focus is on privacy only. Fig. \ref{fig:constant_VaryingPred_targetVariance}  shows that increasing $H_F$ induces a smaller variance on the target load. 

Fig. \ref{fig:PVconstant_VaryingPast_VarianceCompare}, which shows $\mathcal{P}$, $\mathcal{C}$ and $\mathcal{V}$ with respect to $H_P$, exhibits similar behaviors to those in Fig. \ref{fig:PVconstant_VaryingPred_CostCompare}, with some notable differences. The $y$-axis ranges are more limited here, confirming that the knowledge of past consumption is less critical for the EMP compared to the knowledge of future consumption and costs. This explains the far smaller reduction in cost in Fig. \ref{fig:PVconstant_VaryingPast_Cost}, as compared to Fig. \ref{fig:constant_VaryingPred_Cost}, and the increase in the information leakage when $\alpha=1$ in Fig. \ref{fig:PVconstant_VaryingPast_Leakage}. Fig. \ref{fig:PVconstant_VaryingPast_TargetVarianceCompare} shows that the target variance is higher and more variable when $\alpha=1$; however, this corresponds to the case in which the grid load is closer to the target load, i.e., the most private scenario according to our original privacy measure. This contradiction of the two privacy indicators shows that evaluating the variance of the target load does not fully reflect the level of privacy achieved, as defined in Eq. (\ref{eq:privacy}). 

\subsection{Alternative Privacy Measures}

\begin{figure}[!t]
\begin{subfigure}[t]{.5\columnwidth}
\centering
\includegraphics[width=1\columnwidth]{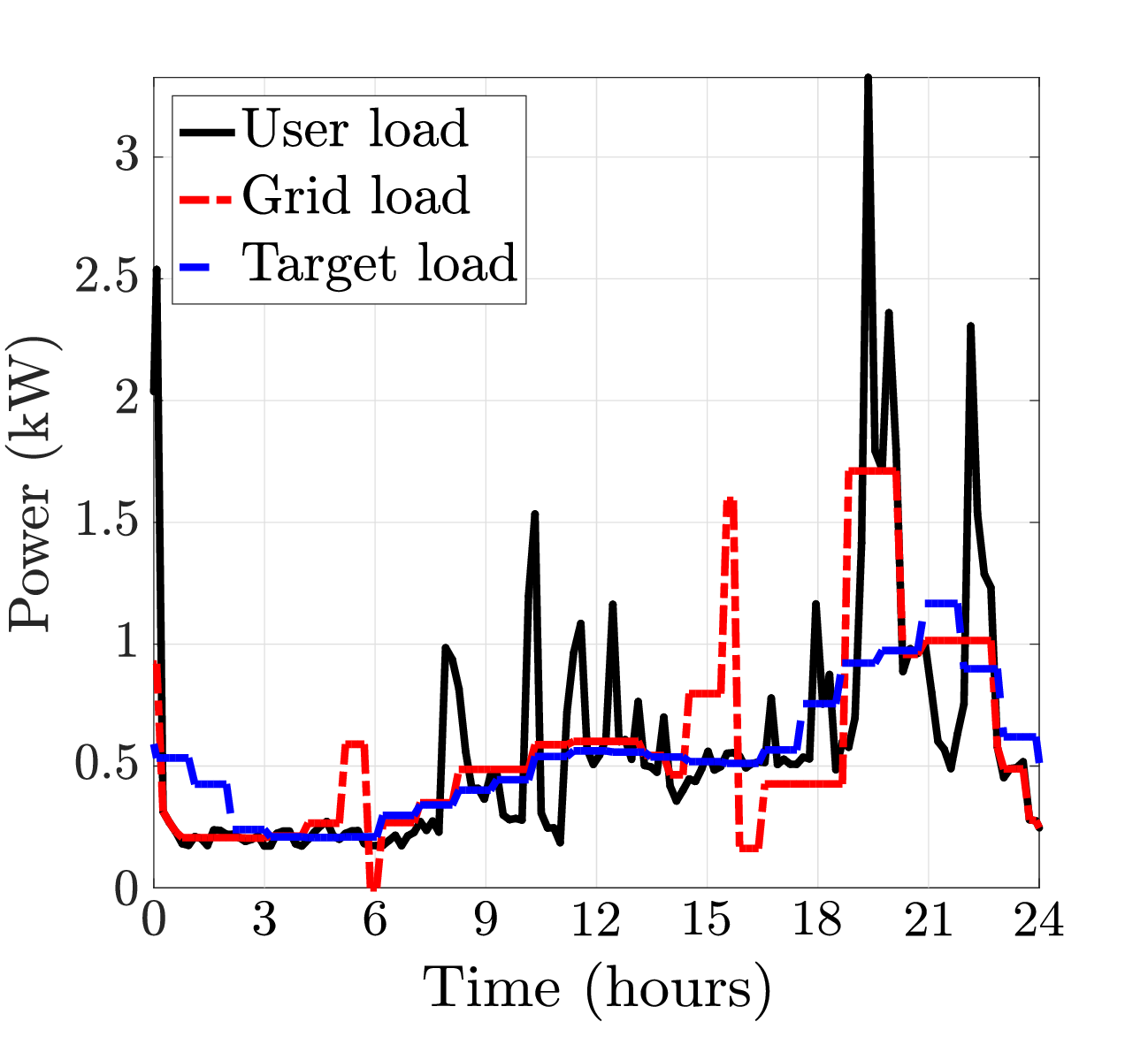}
\caption{Constant target.}
\label{fig:constantShort_Profile}
\end{subfigure}\hfill
\begin{subfigure}[t]{.5\columnwidth}
\centering
\includegraphics[width=1\columnwidth]{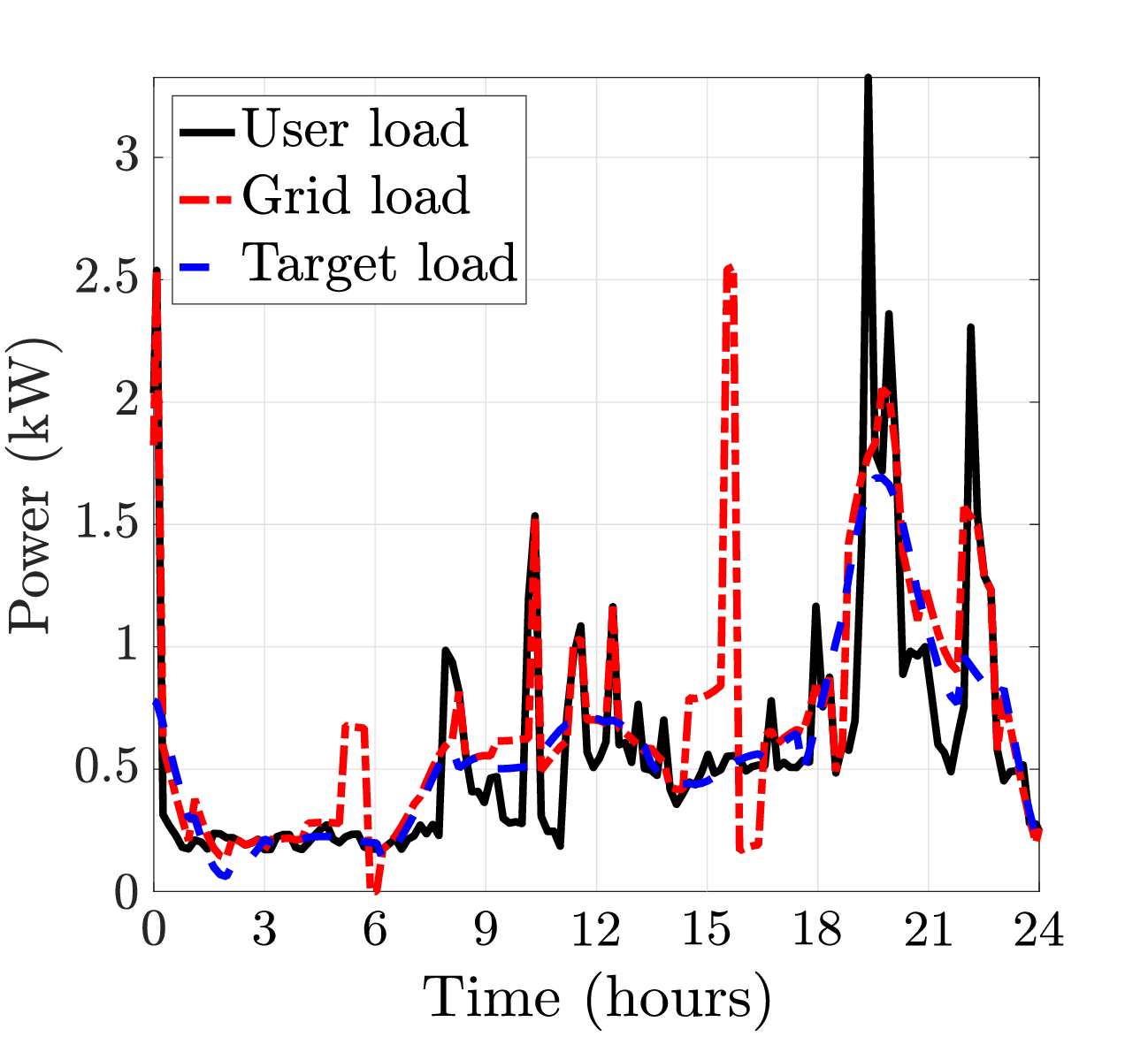}
\caption{Filtered target.}
\label{fig:constantSub_Profile}
\end{subfigure}
\caption{Practical EMP for $\alpha=0.5$, $H_F=H_P=2$h, $T_S=1$h, and cut-off frequency set to $0.1$mHz.}
\label{fig:ConstantShortVSSub}
\end{figure}

\begin{figure*}[!t]
\begin{subfigure}[t]{.33\textwidth}
\centering
\includegraphics[width=1\columnwidth]{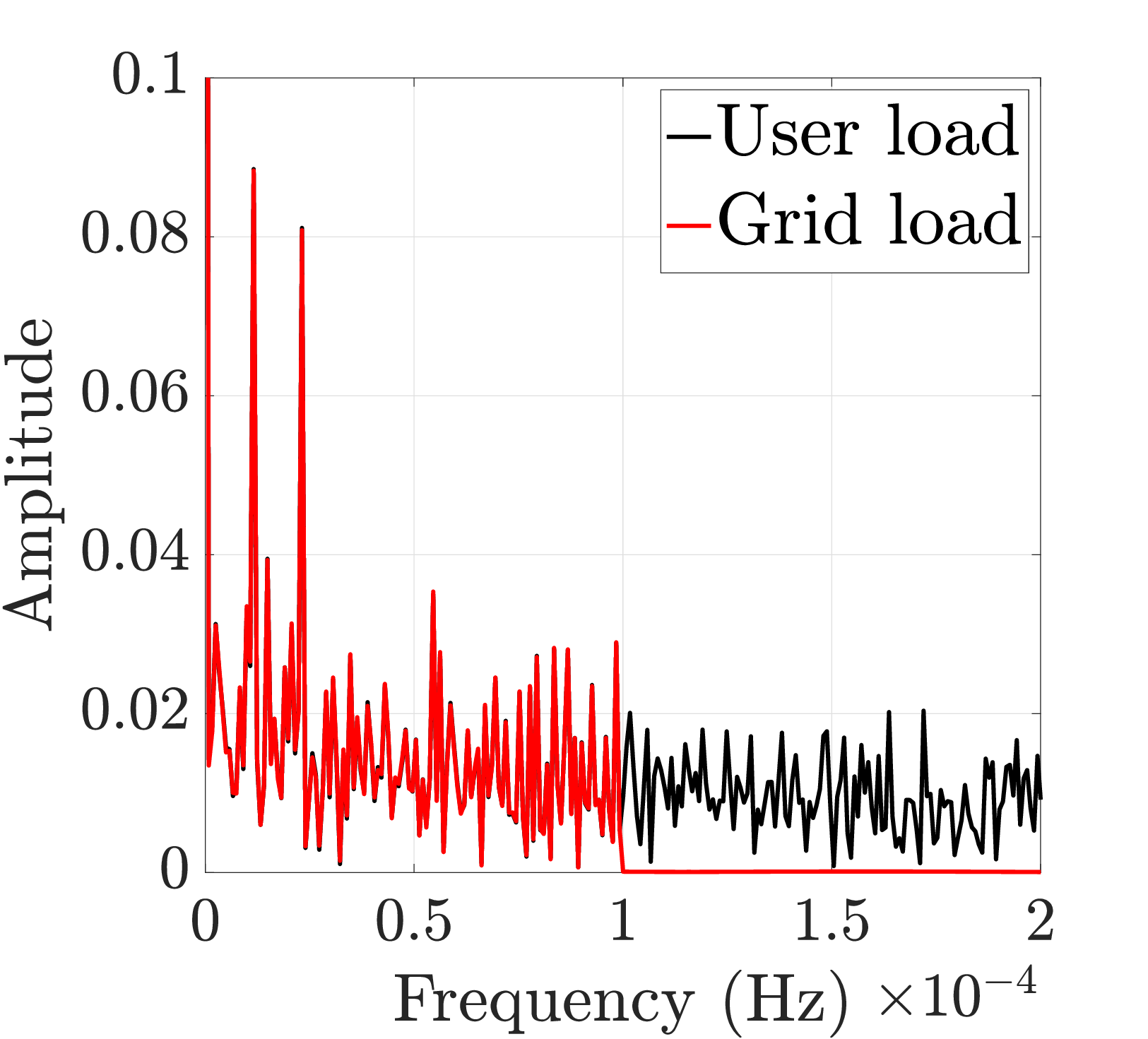}
\caption{LHM.}
\label{fig:filterLong_Spectra}
\end{subfigure}\hfill
\begin{subfigure}[t]{.33\textwidth}
\includegraphics[width=1\columnwidth]{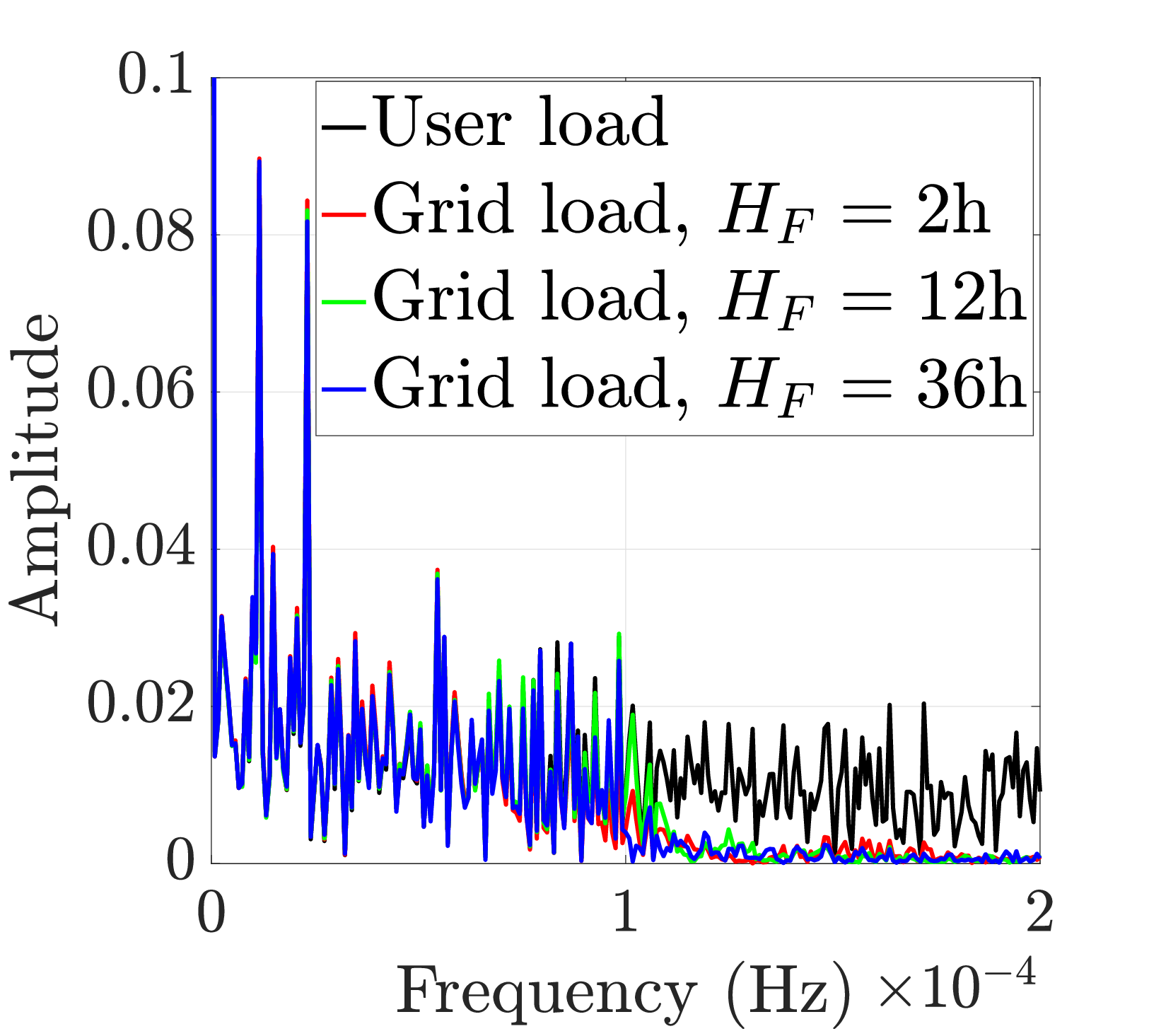}
\caption{SHM.}
\label{fig:filterShort_Spectra3}
\end{subfigure}\hfill
\begin{subfigure}[t]{.33\textwidth}
\includegraphics[width=1\columnwidth]{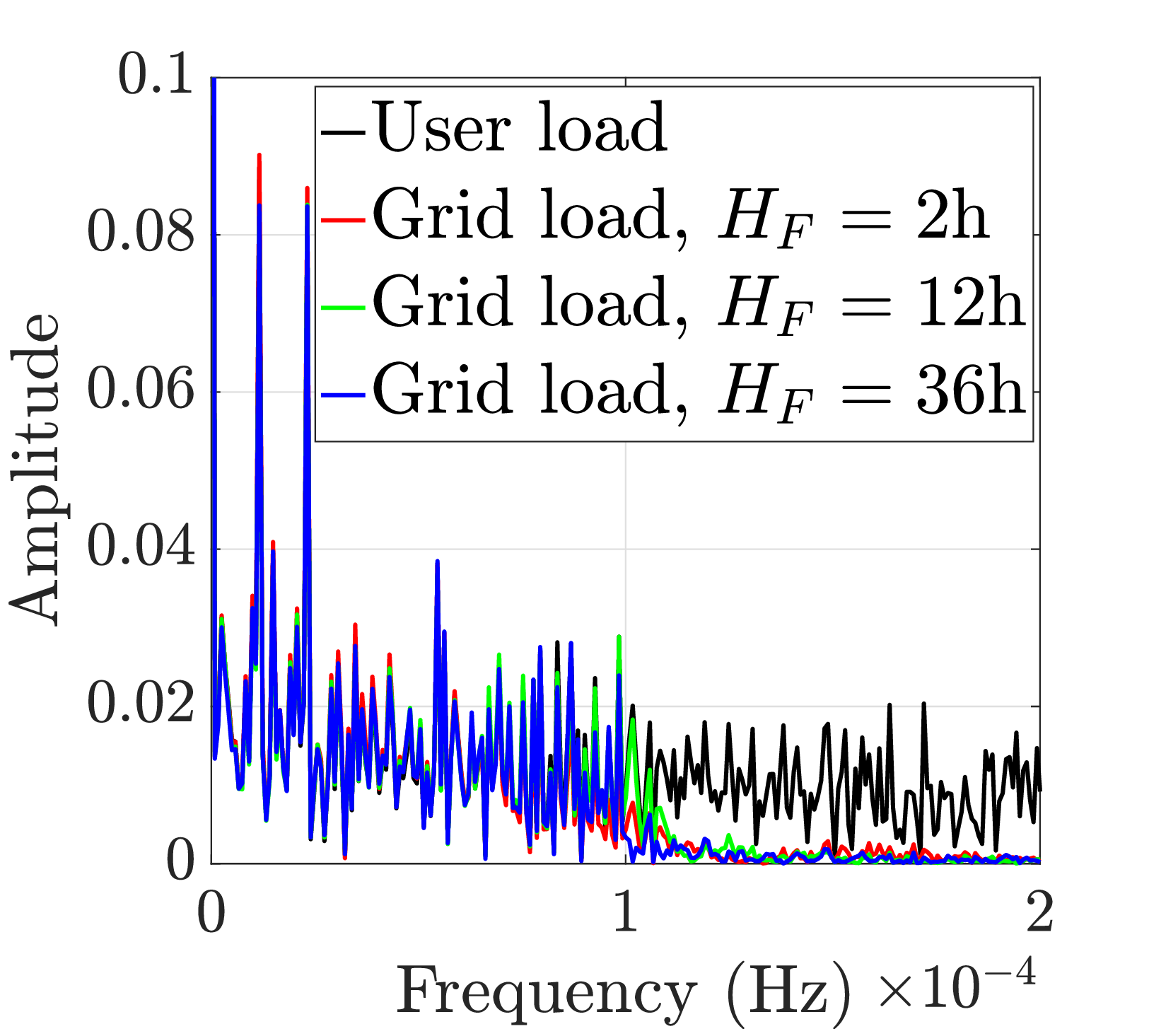}\caption{Practical EMP.}
\label{fig:filterSub_Spectra3}
\end{subfigure}
\caption{Filtered  target load scenario with cut-off frequency of $0.1$mHz, $\alpha=1$, and no energy selling. For the SHM and the practical EMP we set $H_P=2$h and $T_S=1$h.}
\label{fig:filterLongVSShortVSSub_Filter_Spectra05}
\end{figure*}

As opposed to the LHM studied in \cite{Tan:2017TIFS}, where $W_t$ is fixed throughout the operation time, here $W_t$ is allowed to vary over time. Therefore, the squared distance between $G_t$ and $W_t$ may not be sufficient as a privacy measure on its own. Accordingly, we consider alternative measures of privacy to see the impact of the proposed model predictive control framework on those measures. One of the objectives of privacy-preserving algorithms for SM data is to mask the difference between successive power measurements, called \textit{features}, which non-intrusive appliance load monitoring algorithms exploit to identify appliances' switch-on/off events \cite{Prudenzi:2002}. Thus, it is possible to evaluate an EMP's performance against such algorithms by computing the number of features present in the grid load \cite{Yang:2015TSG}. We classify as features those differences that are larger or equal to $50$ W, which represent a typical household electricity consumption of lights. Fig. \ref{fig:constant_VaryingSummary} shows the number of features with respect to $H_F$ and $H_P$. A larger $H_F$ leads to a reduction in the number of features in the grid load (Fig. \ref{fig:constant_VaryingPred_features}); however, $H_P$ does not seem to have any influence on this (Fig. \ref{fig:constant_VaryingPast_features}).

Another way of assessing the performance of privacy-preserving algorithms is by analyzing the power spectrum of the resulting grid load. In fact, the higher-frequency components of the grid load spectrum correspond typically to more sensitive information about a user's energy consumption \cite{Engel:2017}. Fig. \ref{fig:constant_Spectra} shows the grid load spectra corresponding to using different values of $H_F$ and $H_P$. Larger values of $H_F$ lead to better suppression of higher-frequency components when $\alpha=1$ (Fig. \ref{fig:PVconstant_VaryingPred_Spectra1}), whereas for $\alpha=0$ even additional high-frequency components are introduced  (Fig. \ref{fig:PVconstant_VaryingPred_Spectra0selling}). When $\alpha=1$, increasing $H_P$ also attenuates the higher-frequency components (Fig. \ref{fig:PVconstant_VaryingPast_Spectra1_Sellingc}), but less markedly compared to increasing $H_F$. As the spectral analysis of the grid load better captures the information leaked, in the following section we consider a privacy-preserving approach whose aim is to filter out directly the higher-frequency components of the user load. 

\section{Target Load as Filtered User Load}\label{sec:filterTargetLoad}

When the target load is set to a constant value, one can consider this as the DC component of the Fourier transform of the user load profile. If the grid load can be maintained at the average value of the user load at all times, this is equivalent to filtering out all the positive frequency components of the user load profile. However, as shown in Section \ref{sec:constantTargetLoad}, this is not always possible due to the RB capacity and power constraints, and the information leakage is measured as average squared error distance from this constant DC component. In this section, we consider a more general target load profile, obtained by low-pass filtering the user load, which is equivalent to removing only the high-frequency variations. The motivation for this is two-fold: Firstly, the EMU is able to better approximate the target load profile by keeping the low-frequency components; and secondly, the high-frequency components are the ones that leak more information about user behavior. Low-frequency devices are those that typically have continuous periodic operation, e.g., the fridge, and are not particularly privacy sensitive. We would like to remark that, differently from the previous section, here $W$ is not an optimization variable but it is determined based only on the user load. The optimization problem is expressed as
\begin{equation}\label{eq:FilterShort}
\min_{G_t^{\overline{t+H_F}}}    \alpha  \sum_{\tau=t}^{\overline{t+H_F}} (G_{\tau} - W_{\tau})^2  + (1-\alpha)\sum_{\tau=t}^{\overline{t+H_F}} G_{\tau} C_{\tau},
\end{equation}
where $W_t, W_{t+1}, \ldots, W_{\overline{t+H_F}}$ are obtained as low-pass filtered versions of the user load, subject to the same constraints of the constant target scenario, i.e., (\ref{eq:batteryConstraint})-(\ref{eq:energySatisfied}) and (\ref{eq:peakPowerCharging})-(\ref{eq:peakPowerDischarging}). The target load at time $t$, $W_t$, is computed as follows. The EMU selects the only available user load, i.e., that one within $[\overline{t-H_P},\overline{t+H_F}]$, and computes its spectral representation by means of the discrete Fourier transform. Then, a low-pass filter with a predefined cut-off frequency is applied. Finally, the inverse transform provides the target load profile $W_t$. We note that, although $H_P$ does not appear explicitly in (\ref{eq:FilterShort}), the target load computed at time $t$ is determined by low-pass filtering the user load within the time window $[\overline{t-H_P},\overline{t+H_F}]$ to prevent the target load from varying dramatically over different TSs. When $\alpha=0$, Eq. (\ref{eq:FilterShort}) reduces to the linear program of the previous section. The optimal solutions to (\ref{eq:FilterShort}) can be characterized by following the same steps of Section \ref{sec:constantTargetLoad}, apart from $W_t$, which here is not an optimization variable. The optimal solutions are given in (\ref{eq:optimalY}) and (\ref{eq:optimalYNoSel}) for the scenarios where selling energy is allowed and not allowed, respectively. Fig. \ref{fig:filterLongVSShort_Filter} compares the SHM and the LHM, showing that SHM generates profiles that are smooth and similar to that of the LHM, despite relying only the knowledge of $2$ hours of future electricity consumption.

\begin{figure}[!t]
\centering
\includegraphics[width=0.9\columnwidth]{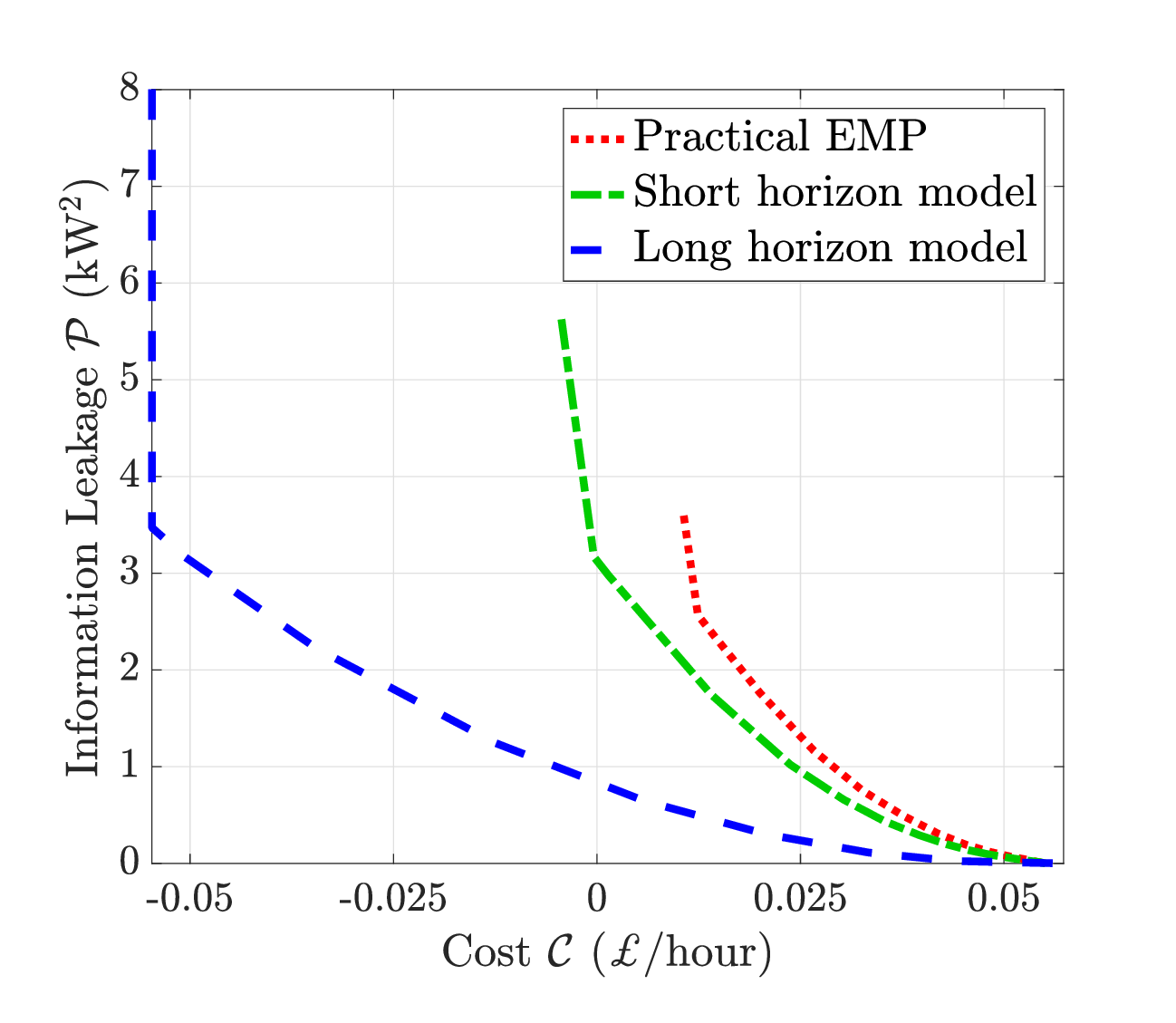}
\caption{The privacy-cost trade-off, for $H_P=H_F=2$h, $T_S=1$h, energy selling allowed, and constant target load for the SHM, LHM and the practical EMP.}
\label{fig:tradeoff}
\end{figure}

\section{A More Practical EMP}\label{sec:suboptimal}

In the previous sections it is assumed that the EMU solves the optimization problem at each TS. However, in practice it may not be feasible to obtain the future predictions at each TS, and may be impractical to compute the target profile so often. Thus, in this section we consider a more practical EMP where the optimization problem is solved once every $T_S$ TSs. The optimization problems at time $t$ are still given by Eqs. (\ref{eq:ConstantShort}) and (\ref{eq:FilterShort}) for the constant and filtered target load scenarios, respectively, such that the sequences $G_{t}^{t+T_S}$ and $W_{t}^{t+T_S}$ are obtained at time $t$ on the basis of the available information for TSs $[t-H_P,t+H_F]$, where $H_F \geq T_S$. 

Fig. \ref{fig:ConstantShortVSSub} shows that the practical EMP for a constant and a filtered target load creates piecewise target and grid loads, similar to the piecewise target load profile approach \cite{Giaconi:2017SGC}. Due to the discontinuities introduced in the grid load profile, spikes at high frequencies may appear in the spectrum of the grid load produced by this strategy, leading to a higher privacy loss. When $\alpha=1$, Fig. \ref{fig:filterLongVSShortVSSub_Filter_Spectra05} shows that the practical EMP reaches virtually the same performance of the SHM, despite computing the grid load six times less often than the LHM.

Finally, in Fig. \ref{fig:tradeoff} we present the privacy-cost trade-offs for the various scenarios we have discussed when energy selling is allowed. This figure clearly highlights the increasing loss in performance due to the decreasing amount of information available to the EMU when moving from the LHM to the SHM and to the more practical EMP.

\section{Conclusions} \label{sec:conclusion}

We have studied the joint optimization of privacy and cost for an SM system equipped with an RB. Privacy is measured via the mean squared-error between the SM measurements and a target load profile, which is set to be either a constant function or a low-pass filtered version of the user load. We assume that only partial information about the user's future electricity consumption and electricity cost is known to the EMU, and we cast the joint privacy and cost optimization as a model predictive control problem. The scenario in which the user is allowed to sell excess energy to the UP is studied, which is shown to achieve a better privacy-cost trade-off. The optimal solutions for the constant and filtered target load profiles have been characterized, highlighting their water-filling interpretation. The privacy-cost trade-off has been characterized for the various scenarios, and detailed numerical simulations and alternative privacy measures have been presented. As a future extension of this study, one can consider a generalization of the SHM and LHM models by introducing errors in the predictions of the future energy consumption profile at the EMU. It is reasonable to assume that the prediction error will increase gradually for more distant time instants in the future.

\section{Acknowledgments}\label{sec11}

G. Giaconi, currently at Ofcom, carried out the research described in this paper while working towards his PhD at Imperial College London. The work of G. Giaconi for this paper was carried out in his personal capacity, and the views expressed here are his own and do not reflect those of his current employer. Giulio Giaconi acknowledges the Engineering and Physical Sciences Research Council (EPSRC) of the UK for funding his PhD studies (award reference \#1507704). The work of H. V. Poor was supported in part by the U.S. National Science Foundation under Grant  ECCS-1824710.

\bibliographystyle{IEEEtran}
\bibliography{bib_IET}

\end{document}